\documentclass[usegraphicx]{mn2e}
\usepackage{amsmath,amssymb}
\usepackage{times}
\usepackage{subfigure}
\usepackage{epsfig}
\usepackage{graphics}

\renewcommand{\descriptionlabel}[1]%
  {\hspace{\labelsep}\textbf{#1}}

\title[Variable star census in NGC 6333]
     {A detailed census of variable stars in the globular cluster NGC~6333 (M9) from 
CCD differential photometry\thanks{Based on observations collected with the 2.0m
telescope at the Indian
Astrophysical Observatory, Hanle, India, and with the Danish 1.54m telescope at the
ESO La Silla Observatory in Chile.}}

\author[A. Arellano Ferro et al.]
{A. Arellano Ferro$^{1}$\thanks{E-mail:armando@astro.unam.mx}, 
D.M. Bramich$^{2}$,
R. Figuera Jaimes$^{2,3}$,
Sunetra Giridhar$^{4}$,
N. Kains$^{2}$, \and
K. Kuppuswamy$^{4}$,
U. G. J{\o}rgensen$^{5,14}$\and
and \and
K.A. Alsubai$^{6}$,
J. M. Andersen$^{7,14}$,
V. Bozza$^{8,9}$,
P. Browne$^{3}$,
S. Calchi Novati$^{8,10}$\and 
Y. Damerdji$^{11}$, 
C. Diehl$^{12,25}$, 
M. Dominik$^{3}$,
S. Dreizler$^{13}$,
A. Elyiv$^{11,27}$, 
E. Giannini$^{12}$,\and
K. Harps{\o}e$^{5,14}$,
F.V. Hessman$^{13}$,
T.C. Hinse$^{15,5}$,
M. Hundertmark$^{3}$,
D. Juncher$^{5,14}$, \and
E. Kerins$^{16}$,
H. Korhonen$^{5,14}$,
C. Liebig $^{3}$,
L. Mancini$^{17}$,
M. Mathiasen$^{5}$,
M.T. Penny $^{18}$,\and
M. Rabus$^{19}$,
S. Rahvar$^{20,21}$,
D. Ricci$^{11,26}$,
G. Scarpetta$^{8,22}$,
J. Skottfelt$^{5,14}$,\and
C. Snodgrass$^{23}$,
J. Southworth$^{24}$,
J. Surdej$^{11}$, 
J. Tregloan-Reed$^{24}$,
C. Vilela$^{24}$,\and
O. Wertz$^{11}$
\large{(The MiNDSTEp consortium)}\and
\large{\it{(Affiliations can be found after the references)}}
}

\begin{document} 

\date{Accepted 2013 June 13. Received 2013 June 9; in original form 2013 May 19}

\pagerange{\pageref{1}--\pageref{15}} \pubyear{2013}

\maketitle 

\label{firstpage}

\begin{abstract}
We report CCD $V$ and $I$ time-series photometry of the globular cluster NGC 6333
(M9). The technique of difference image analysis has been used, which enables
photometric precision better than 0.05 mag for stars brighter than $V \sim 19.0$
mag.,
even in the crowded central regions of the cluster. 
The high photometric precision has resulted in the discovery of two new RRc stars, 
three eclipsing binaries, seven long-term variables and one field RRab
star behind the cluster. A detailed
identification chart and equatorial coordinates are given for all the variable stars
in the field of our images of the cluster. 
Our data together with literature $V$-data obtained in 1994 and 1995 allowed us to
refine considerably the periods for all RR Lyrae stars. The nature of the new
variables is discussed. We argue that variable V12 is a cluster member and an 
Anomalous Cepheid. Secular period variations, double mode pulsations and/or the
Blazhko-like modulations in some RRc variables are addressed.
Through the light curve Fourier decomposition of 12 RR Lyrae stars we have
calculated a mean metallicity of 
[Fe/H]$_{\rm ZW}$=$-1.70 \pm 0.01{\rm(statistical)} \pm 0.14{\rm(systematic)}$ 
or [Fe/H]$_{UVES}=-1.67 \pm 0.01{\rm(statistical)} \pm 0.19{\rm(systematic)}$.
Absolute magnitudes, radii and masses are also estimated for the RR Lyrae stars.

A detailed search for SX Phe stars in the Blue Straggler region was conducted but
none were discovered. If SX Phe exist in the cluster then their amplitudes must be
smaller than the detection limit of our photometry.

The CMD has been corrected for heavy differential reddening using the detailed
extinction map of the cluster of Alonso-Garc\'ia et al. (2012). This has allowed us
to set the mean cluster distance from two independent estimates; from the RRab and RRc
absolute magnitudes, we find $8.04\pm 0.19$ kpc and $7.88\pm0.30$ kpc respectively.

\end{abstract}
      

\section{Introduction}

In the last thirty years the variable stars in the globular cluster NGC~6333 (M9
or C1716-184 in the IAU nomenclature) ($\alpha = 17^{\mbox{\scriptsize h}}
19^{\mbox{\scriptsize m}}
11.8^{\mbox{\scriptsize s}}$, $\delta = -18\degr 30\arcmin 58.5\arcsec$, J2000;
$l = 5.54\degr$, $b = +10.71\degr$) have
been the subject of some analyses based on photographic and CCD time-series photometry
(Clement, Ip \& Robert 1984; Clement \& Walker 1991; Clement \& Shelton 1996; 1999).
The 2012 update for NGC~6333 in the Clement et al. (2001) Catalogue of Variable Stars
in Globular Clusters (CVSGC) lists 21 known variable stars; 9 RRab, 9 RRc, , 1 long 
period variable (V8), 1 Pop II cepheid (V12) and 1 eclipsing binary (V21). This
makes the cluster
attractive for a Fourier decomposition analysis of the RR Lyrae star light curves with
the aim of calculating their physical parameters from semi-empirical calibrations.
Furthermore,
this cluster has a crowded central region where it is difficult to perform conventional
point-spread-function (PSF) fitting photometry. The application of difference image
analysis (DIA) to image data for this cluster for the first time therefore opens up the
possibility of new variable star discoveries.

Recently our team has performed CCD photometry of several globular clusters 
by employing the DIA technique to produce precise
time-series photometry of individual stars down to $V \sim$ 19.5 mag. The DIA
photometry has proven to be a very useful tool in obtaining high quality light curves
of known variables, and for discovering and classifying new variables (e.g. Arellano
Ferro et al. 2011; Bramich et al. 2011; Kains et al. 2012; Figuera Jaimes et al. 2013;
and references therein), where previous CCD photometric studies have not detected
stellar variability, particularly in the crowded central regions of the clusters.
Thus, in the present paper we report the analysis of new time-series photometry of
NGC 6333 in the $V$ and $I$ filters.
In $\S$ 2 we describe the observations and data reductions. In $\S$ 3 the problem of the
differential reddening in the cluster field-of-view (FOV) is addressed and the approach
we used to correct it is described. $\S$ 4 contains a
detailed discussion on the approach to the identification of new variables
and their classification. In $\S$ 5 we apply Fourier light curve decomposition
to some of the RR Lyrae stars and calculate their metallicity and absolute magnitude.
Given the differential reddening correction, the accuracy in the cluster distance 
determination is highlighted. In $\S$ 6 we discuss the $A_V-\log P$ relation for the
RR Lyrae stars and  the Oosterhoff type of the cluster. In $\S$ 7 we summarize our
results.

\section{Observations and Reductions}
\label{sec:Observations}

\subsection{Observations}

The observations employed in the present work were performed using the Johnson $V$ 
and $I$ filters on 15 nights during 2010-2012 at two different observatories.
The 2.0m telescope of the Indian Astronomical Observatory (IAO), Hanle, India,
located at 4500m above sea level, was used to obtain 212 and 171 epochs in the
$V$ and $I$ filters, respectively. The detector was a Thompson CCD of 2048$\times$2048
pixels with a pixel scale of 0.296 arcsec/pix translating to a FOV of approximately
10.1$\times$10.1~arcmin$^2$. Also, the Danish Faint Object Spectrograph and Camera (DFOSC)
at the Danish 1.54m telescope at La Silla, Chile, was used to collect 118 epochs
in the $V$ filter. DFOSC has a 2147$\times$2101 pixel Loral CCD with a pixel scale
of 0.396~arcsec/pix and a FOV of $\sim$14.2$\times$13.9~arcmin$^2$.

The log of observations is shown in Table \ref{tab:observations} where the dates,
site, number of frames, exposure times and average nightly seeing are recorded. A total of 
330 epochs in the $V$ filter and 171 in the $I$ filter spanning just over 2 years
are included in this study.

\begin{table*}
\caption{The distribution of observations of NGC 6333 for each filter, where the
columns $N_{V}$ and $N_{I}$ represent the number of images taken with the $V$ and $I$
filters respectively. We also provide the exposure time, or range of exposure times,
employed during each night for each filter in the columns $t_{V}$ and $t_{I}$ and the
average seeing in the last column.}
\centering
\begin{tabular}{lcccccc}
\hline
Date          & Telescope                      & $N_{V}$ & $t_{V}$ (s) & $N_{I}$ &
$t_{I}$ (s)&Avg seeing (") \\
\hline
20100506 & 2.0m IAO, Hanle, India & 29            & 180-250       & 26            &
40-60 &2.2\\
20110412 & 2.0m IAO, Hanle, India & 4             & 75-125         & 5              &
10-60 &1.5\\
20110413 & 2.0m IAO, Hanle, India & 11           & 70                 & 11           &
8-10 &1.4\\
20110414 & 2.0m IAO, Hanle, India & 14           & 70                 & 14           &
10 &1.4\\
20110610 & 2.0m IAO, Hanle, India & 13            & 70-100         & 13            &
10-15 &1.7\\
20110611 & 2.0m IAO, Hanle, India & 3              & 200-250       & 3              &
10-40 &2.8\\
20110805 & 2.0m IAO, Hanle, India & 13            & 100-200       & 13            &
15-50 &1.8\\
20110806 & 2.0m IAO, Hanle, India & 6              & 90-100         & 6              
& 12-15 &1.9\\
20120515 & 2.0m IAO, Hanle, India & 3              & 20-40          & 3              
& 7-20 &1.9\\
20120516 & 2.0m IAO, Hanle, India & 23            & 10-35          & 24             &
4-35 &1.8\\
20120628 & 2.0m IAO, Hanle, India & 52            & 20-60          & 53             &
3-15 &2.2\\
20120629 & 2.0m IAO, Hanle, India & 41            & 40-100       & ---              &
--- &2.7\\
20120822 & 1.54m Danish, La Silla, Chile & 28 & 100-180    & ---              & ---
&3.1\\
20120824 & 1.54m Danish, La Silla, Chile & 39 & 50-100      & ---               & ---
&1.8\\
20120826 & 1.54m Danish, La Silla, Chile & 51 & 50             & ---               &
---&1.0\\
\hline
Total:   &  & 330     &             &  171    &           &\\
\hline
\end{tabular}
\label{tab:observations}
\end{table*}

\subsection{Difference Image Analysis}

We employed the technique of difference image analysis (DIA) to extract high-precision
photometry for all of the point sources in the images of NGC~6333 and we used the 
{\tt DanDIA}\footnote{{\tt DanDIA} is built from the DanIDL library of IDL routines
available at http://www.danidl.co.uk}
pipeline for the data reduction process (Bramich et al. 2013) which includes an 
algorithm that models the convolution kernel matching the PSF
of a pair of images of the same field as a discrete pixel array (Bramich 2008).

\begin{figure} 
\includegraphics[width=8.0cm,height=8.0cm]{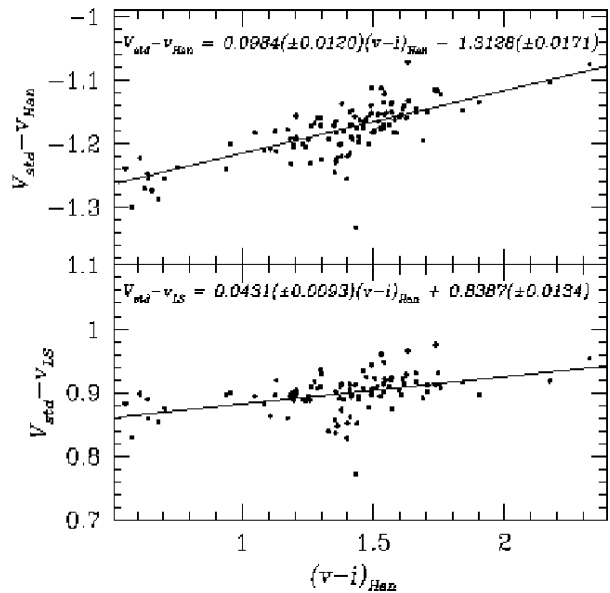}
\caption{Transformation relations between the instrumental and the standard
photometric systems using a set of standard stars in the field of NGC~6333 provided
by Peter Stetson. Top and bottom panels correspond to the observations from Hanle and 
La Silla respectively. The lack of $I$-band standards forced us to leave the Hanle $I$
data in the instrumental system. The $V$ observations from La Silla have been fit with
the Hanle colour $(v-i)_{Han}$ to reveal the colour dependence. See $\S$
\ref{absolute} for a discussion.}
    \label{transV}
\end{figure}

\begin{figure} 
\includegraphics[width=8.0cm,height=8.0cm]{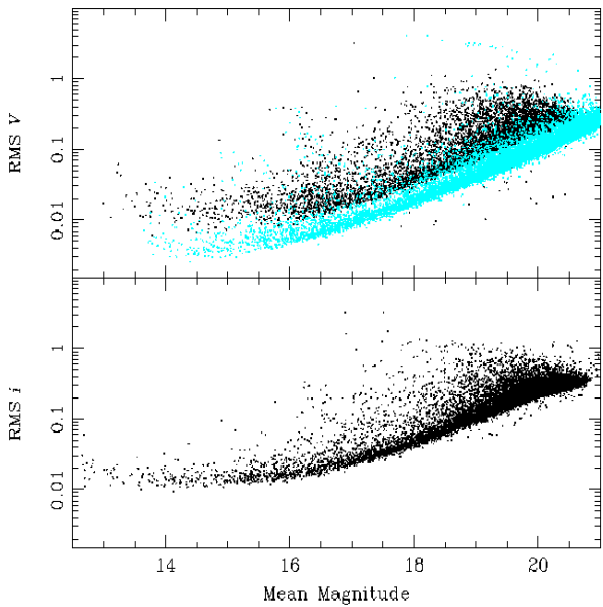}
\caption{The rms magnitude deviations as a function of magnitude. The upper panel corresponds to the $V$ light curves from
Hanle (black dots, 12519 stars) and La Silla  (cyan dots, 21143 stars). The lower
panel corresponds to the $i$ light curves from Hanle (15524 stars).}
\label{rms}
\end{figure}

The {\tt DanDIA} pipeline performs standard overscan bias level and flat field corrections
of the raw images, and creates a reference image for each filter by stacking a set of
registered best-seeing calibrated images. For the La Silla image data, which are slightly
undersampled in the best-seeing images, it was necessary to pre-blur any images with a
seeing of less than 3~pix to force a full-width at half-maximum (FWHM) of the PSF of at
least 3~pix. This is because undersampling can cause problems in determining the kernel
solution matching the PSFs between images.

We constructed three reference images, one in
the $V$ filter for each telescope and one in $I$ for the Hanle data. For each of these
reference images, 5, 10 and 6 calibrated images were stacked with total exposure times of
350, 500 and 50~s in the $V$ (Hanle), $V$ (La Silla) and $I$ (Hanle) filters, respectively,
with PSF FWHMs of $\sim$4.3, $\sim$3.2, and $\sim$3.5~pix, respectively.
In each reference image, we measured the fluxes (referred to as reference fluxes) 
and positions of all PSF-like objects (stars)
by extracting a spatially variable (with a third-degree polynomial) empirical PSF from the 
image and fitting this PSF to each detected object. The detected stars in each image
in the time-series were matched with those detected in the corresponding reference
image,
and a linear transformation was derived which was used to register each image with
the reference image.

For each filter, a sequence of difference images was created by 
subtracting the relevant reference image, convolved with an appropriate spatially
variable kernel, from each registered image. The spatially variable convolution
kernel for each registered image was determined using bilinear interpolation of a
set of kernels that were derived for a uniform 6$\times$6 grid of subregions across the
image.

The differential fluxes for each star
detected in the reference image were measured on each difference image. 
Light curves for each star were constructed by calculating the total 
flux $f_{\mbox{\scriptsize tot}}(t)$ in ADU/s at each epoch $t$ from:
\begin{equation}
f_{\mbox{\scriptsize tot}}(t) = f_{\mbox{\scriptsize ref}} +
\frac{f_{\mbox{\scriptsize diff}}(t)}{p(t)}
\label{eqn:totflux}
\end{equation}
where $f_{\mbox{\scriptsize ref}}$ is the reference flux (ADU/s),
$f_{\mbox{\scriptsize diff}}(t)$ is the differential flux (ADU/s) and
$p(t)$ is the photometric scale factor (the integral of the kernel solution).
Conversion to instrumental magnitudes was achieved using:
\begin{equation}
m_{\mbox{\scriptsize ins}}(t) = 25.0 - 2.5 \log \left[ f_{\mbox{\scriptsize tot}}(t) \right]
\label{eqn:mag}
\end{equation}
where $m_{\mbox{\scriptsize ins}}(t)$ is the instrumental magnitude of the star 
at time $t$. Uncertainties were propagated in the correct analytical fashion.

The above procedure and its caveats have been described in detail in Bramich et al.
(2011) and the interested reader is referred there for the relevant details.

\begin{figure*} 
\includegraphics[width=17.0cm,height=9.0cm]{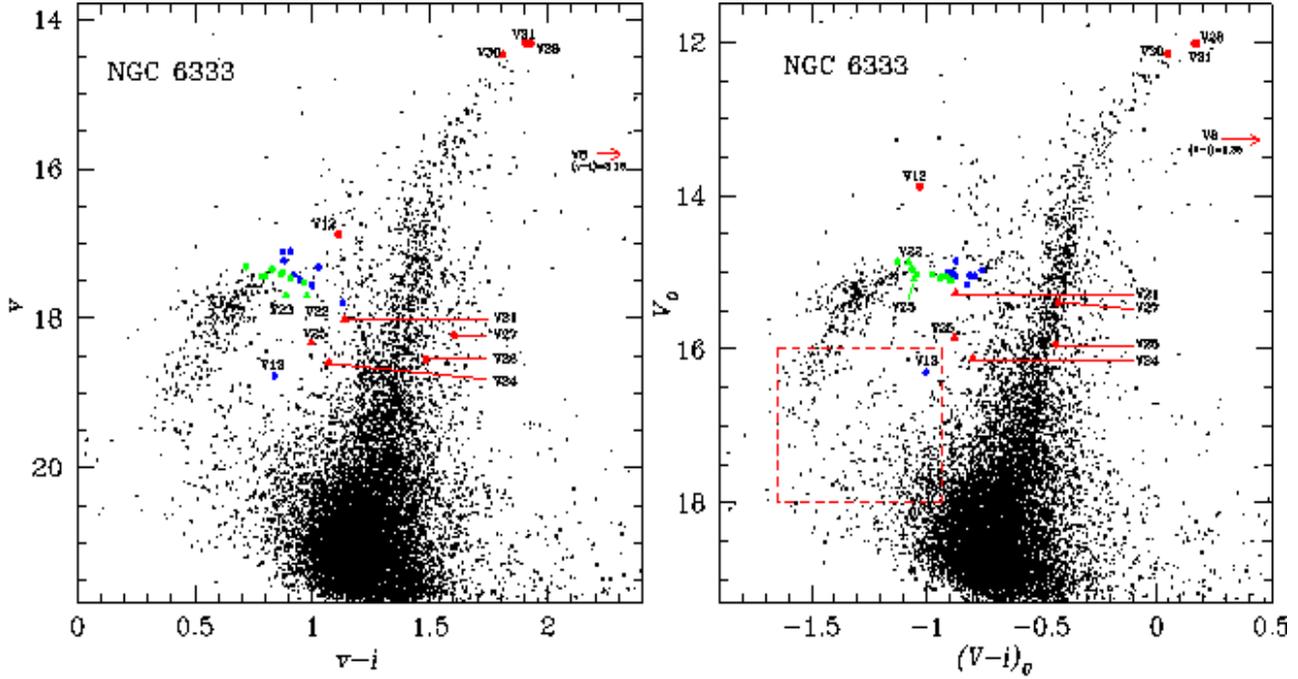}
\caption{Colour-magnitude diagram of NGC~6333 for the Hanle data in the instrumental
system (left) and 
after transforming $v$ into the standard system and correcting for differential
reddening (right) as decribed in the text. The coloured symbols correspond to the
known variables and the new variables
discovered in this paper. The colour coding is as follows: RRab stars - blue circles,
known RRc stars - green circles, new RRc stars (V22, V23) - green triangles, long
period
variables (V8, V26, V27, V28, V30, V31) and
anomalous cepheid (V12) - red circles, eclipsing binaries - red triangles. The region bounded
by red dashed lines is an arbitrarily defined Blue Straggler region. New variables or
interesting stars are discussed in $\S$ \ref{sec:IND_STARS} are labelled. The
variables 
V29, V32, V33 and V34 are not plotted because they are either saturated in the $I$
images, or outside the FOV of the Hanle data.}
\label{CMD}
\end{figure*}

\begin{table*}
\caption{Time-series $V$ and $i$ photometry for all the confirmed variables in our 
field of view. The telescope employed is coded in column 2 (2.0H=2m telescope in
Hanle; 1.5LS=1.54m telescope in La Silla). The standard $M_{\mbox{\scriptsize std}}$ and
instrumental $m_{\mbox{\scriptsize ins}}$ magnitudes are listed in columns 5 and 6,
respectively, corresponding to the variable star in column 1. Filter and epoch of
mid-exposure are listed in columns 3 and 4, respectively. The uncertainty on
$m_{\mbox{\scriptsize ins}}$ is listed in column 7, which also corresponds to the
uncertainty on $M_{\mbox{\scriptsize std}}$. For completeness, we also list the quantities
$f_{\mbox{\scriptsize ref}}$, $f_{\mbox{\scriptsize diff}}$ and $p$ from Equation~\ref{eqn:totflux}
in columns 8, 10 and 12, along with the uncertainties $\sigma_{\mbox{\scriptsize ref}}$
and $\sigma_{\mbox{\scriptsize diff}}$ in columns 9 and 11. This is an extract from the full
table, which is available with the electronic version of the article (see Supporting Information).
         }
\centering
\begin{tabular}{cccccccccccc}
\hline
Variable &Telescope &Filter & HJD & $M_{\mbox{\scriptsize std}}$ &
$m_{\mbox{\scriptsize ins}}$
& $\sigma_{m}$ & $f_{\mbox{\scriptsize ref}}$ & $\sigma_{\mbox{\scriptsize ref}}$ &
$f_{\mbox{\scriptsize diff}}$ &
$\sigma_{\mbox{\scriptsize diff}}$ & $p$ \\
Star ID  &  &      & (d) & (mag)                        & (mag)                       
& (mag)        & (ADU s$^{-1}$)               & (ADU s$^{-1}$)                    &
(ADU s$^{-1}$)                &
(ADU s$^{-1}$)                     &     \\
\hline
V1 &2.0H& $V$ & 2455323.25294    & 16.512 & 17.735 & 0.004  & 843.914 & 3.864 
&-28.802 & 2.353 & 0.7536 \\
V1 &2.0H& $V$ & 2455323.26683    & 16.520 & 17.742 & 0.004  & 843.914 & 3.864  &-34.345 & 2.301 & 0.7816  \\
\vdots   & \vdots & \vdots  & \vdots & \vdots & \vdots & \vdots   & \vdots & \vdots  & \vdots & \vdots \\
V1 &2.0H & $I$ &2455323.27116 &0.000 &16.632& 0.007& 2073.761 &12.792& +115.682&
10.584 &0.7681\\
V1 &2.0H & $I$ &2455323.29722 &0.000 &16.636& 0.006& 2073.761 &12.792& +110.797&
10.198 &0.7825\\
\vdots   & \vdots & \vdots  & \vdots & \vdots & \vdots & \vdots   & \vdots & \vdots  & \vdots & \vdots \\
V1 &1.5LS &$V$& 2456162.62183 &16.488 &15.615& 0.004 &9079.038& 11.294& -3191.886& 21.994 &0.9373\\
V1 &1.5LS &$V$& 2456162.62329 &16.486 &15.613& 0.004 &9079.038& 11.294& -3193.007& 21.104 &0.9406\\
\hline
\end{tabular}
\label{tab:vi_phot}
\end{table*}

\subsection{Photometric Calibrations}

\subsubsection{Relative}
\label{sec:rel}

All photometric data suffer from systematic errors to some level. Sometimes they may be severe enough
to be mistaken for bona fide variability in light curves (e.g. Safonova \& Stalin 2011). However,
multiple observations of a set of objects at different epochs, such as time-series photometry,
may be used to investigate, and possibly correct, these systematic errors (see for example
Honeycutt 1992). This process is a relative self-calibration of the photometry, which is being
performed as a standard post-processing step for large-scale surveys (e.g. Padmanabhan et al. 2008;
Regnault et al. 2009; etc.).

We apply the methodology developed in Bramich \& Freudling (2012) to solve for the magnitude offsets
$Z_{k}$ that should be applied to each photometric measurement from the image $k$. In terms of DIA,
this translates into a correction (to first order) for the systematic error introduced into the photometry
from an image due to an error in the fitted value of the photometric scale factor $p$. We found that
for Hanle images in either filter the magnitude offsets that we derive are of the order of $\sim$10~mmag
with a handful of worse cases reaching $\sim$30~mmag. For the La Silla data, the
magnitude offsets that we derive
are of the order of $\sim$1-5~mmag. Applying these magnitude offsets to our DIA photometry notably improves
the light curve quality, especially for the brighter stars.

\subsubsection{Absolute}
\label{absolute}

Standard stars in the field of NGC~6333 are not included in the online collection of
Stetson (2000)\footnote{http://www3.cadc-ccda.hia-iha.nrc-cnrc.gc.ca/community/STETSON/standards}.
However, Prof. Stetson has kindly provided us with a set of preliminary standard 
stars which we have used to transform instrumental $v$ magnitudes into the standard
$V$ system. The lack of equivalent values in the $I$ filter forced us to leave our
observations for this filter in the instrumental system. 

The standard minus the instrumental magnitudes show mild dependences on the colour,
as can be seen in Fig.\ref{transV}. The transformations are of the form

\begin{equation}
V_{std}= v_{Han} +0.0984(\pm0.0120)(v-i)_{Han} - 1.3128(\pm0.0171),
\label{eq:transHan}
\end{equation}

\begin{equation}
V_{std}= v_{LS} +0.0431(\pm0.0093)(v-i)_{Han} + 0.8387(\pm0.0134).
\label{eq:transLHa}
\end{equation}

Due to the lack of observations in the $I$-band for La Silla, Eq. \ref{eq:transLHa}
was fit using the Hanle colour $(v-i)_{Han}$. For the La Silla data, we have
adopted
$(v-i)_{Han}$=0.8 mag which corresponds approximately the centre of RR Lyrae
horizontal
branch (HB), hence 

\begin{equation}
V_{std}= v_{LS} + 0.8731(\pm0.0163),
\label{eq:transLS}
\end{equation}

\noindent
which we have used to tranform the instrumental into the standard magnitudes
for La Silla. Given the instrumental colour range of RR Lyrae stars, 0.60-1.0, this
practice produces standard $V$ magnitudes consistent with the zero point
uncertainties of the
above equations. For much redder variables, like those at the tip of the red giant
branch (RGB), the standard $V$ magnitudes for the La Silla data may be off by as much
as 0.025 mag. Due to saturation in the $I$-band images, variable V29 has no
$v-i$ value and we have also adopted $(v-i)_{Han}$=0.8 to calculate its standard $V$
magnitudes.

All of our $Vi$ photometry for the variable stars in the field of the La Silla images
of NGC 6333 is provided in Table \ref{tab:vi_phot}. Only a small portion of this
table is given in the printed version of this paper while the full table is available
in electronic form.

Fig. \ref{rms} shows the rms magnitude deviation in our $V$ and $i$ light curves, 
after the relative photometric calibration of Section~\ref{sec:rel}, as a function of
mean magnitude. We achieve rms scatter at the bright end of $\sim$10-20~mmag in both
the $V$ and $I$ filters for the Hanle data, while we achieve an even better rms for
all magnitudes for the La Silla data with an rms at the bright end of $\sim$3-5~mmag.
We believe that the La Silla data, which come from the smaller of the two telescopes,
performed significantly better in terms of S/N due to a combination of better seeing
for the majority of images and more stable flat fielding.

\subsection{Astrometry}
\label{sec:astrometry}

A linear astrometric solution was derived for the $V$ filter reference image from
La Silla (which has the larger FOV) by matching $\sim$600 hand-picked stars with the 
UCAC3 star catalogue (Zacharias et al. 2010) using a field overlay in the image
display tool {\tt GAIA} 
(Draper 2000). We achieved a radial RMS scatter in the residuals of
$\sim$0.3~arcsec. The astrometric fit was then used to calculate the J2000.0 
celestial coordinates for all of the confirmed variables in our field of view
(see Table~\ref{variables}). The coordinates correspond to the epoch of the $V$
reference image from La Silla, which pertains to the heliocentric Julian day
2456166.51~d.

\section{Reddening}
\label{sec:reddening}

NGC 6333 is known to have differential reddening (Clement, Ip \& Robert 1984) with a
heavily obscuring cloud to the SW of the cluster which is evident in the cluster
images. Hence, without a proper correction for the
differential reddening effects it would be difficult to tell how much of the
dispersion in the position of the RR Lyrae stars in the CMD is due to reddening and
how much is due to physical and evolutionary effects. 
The large dispersion in the HB and RGB in the
uncorrected CMD shown in the left-hand panel of Fig.~\ref{CMD} is evident, and it is particularly
visible in the distribution of the RR Lyrae stars.

Foreground reddening estimates for NGC~6333 can be found in the literature,
e.g. $E(B-V)=0.34$ mag (Zinn 1985); 0.32-0.37 mag (Reed, Hesser \& Shawl 1988); 0.38 mag (Harris
1996, 2010 edition).

To correct for the differential reddening we have taken advantage of the detailed reddening maps
calculated by Alonso-Garc\'ia et al. (2012) for a group of globular clusters
in the inner Galaxy including NGC~6333. In the map for this cluster,
differential reddenings are
presented for a grid of 27668 coordinates within $\sim$11~arcmin centred in the
field of the cluster and with a spatial resolution of $\sim$3.6~arcsec. For each star in our
Hanle reference images, we averaged the differential reddenings for the four neighbouring
values in the grid and follow Alonso-Garc\'ia et al. in using the absolute extinction zero point of
$E(B-V)=0.43$~mag, estimated by comparing with the map of Schlegel, Finkbeiner \&
Davis (1998), to obtain a reddening. Then we corrected our $V$ and $i$ magnitudes for
each star by adopting a normal extinction $A_V = 3.1E(B-V)$ and the ratio $A_I/A_V=
0.479$ (Cardelli, Clayton, \& Mathis 1989) from which $E(V-I)/E(B-V)=1.616$
follows. The resulting corrected CMD is shown in the right-hand panel of
Fig. \ref{CMD}. The RR Lyrae stars have been plotted using their intensity-weighted
magnitudes calculated by fitting eq. \ref{eq_foufit} ($A_o$) to their light curves.

\begin{figure} 
\includegraphics[width=6.0cm,height=8.5cm]{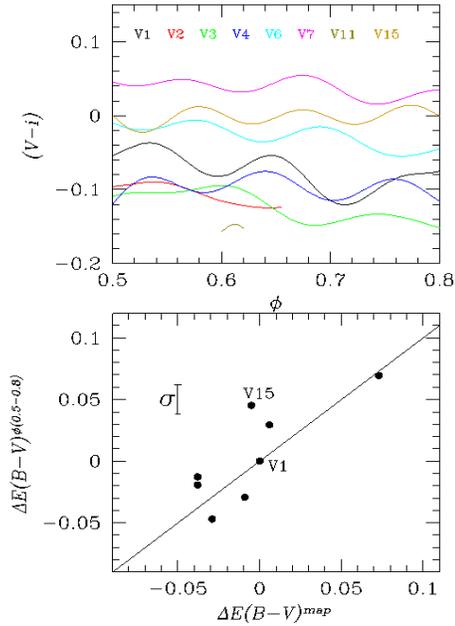}
\caption{The top panel shows the difference between the Fourier fits to the standard 
$V$ and instrumental $i$ magnitudes for the light curves of the RRab stars in the
phase interval 0.5-0.8. The different mean levels of these curves are due to the
differential reddening affecting the cluster. V13 is not included in this analysis
since it is not a cluster member (see Sec.~\ref{sec:IND_STARS}).
The Fourier fits of V2 and V11 are shown only in part since their light curves
are not fully covered by data in the 0.5-0.8 phase range. In the bottom panel we compare 
the differential values $\Delta E(B-V)^{\phi(0.5-0.8)}$ of a given variable relative 
to V1 with $\Delta E(B-V)^{map}$ also relative to V1. The error bar represents the
typical uncertainty in the colour differences.}
\label{fig:red-VI}
\end{figure}

Individual reddenings for the RRab stars may also be calculated from their colour near minimum
light. The method, originally proposed by Sturch (1966), has been further investigated
by Blanco (1992), Mateo et al. (1995), Guldenschuh et al. (2005) and Kunder et al.
(2010). Guldenschuh et al. (2005) concluded that for RRab stars the intrinsic colour
between phases 0.5 and 0.8 is $(V-I)_o^{\phi(0.5-0.8)}=0.58 \pm 0.02$~mag. For the RRab
stars in NGC~6333 we have calculated the mean $V-i$ colour curves as the difference between
the Fourier fits from eq. \ref{eq_foufit} for the standard $V$ and instrumental $i$ light curves.
A plot of the colour curves in the 0.5-0.8 phase interval is shown for each star in the top
panel of Fig. \ref{fig:red-VI}. The different mean levels of the
$(V-i)^{\phi(0.5-0.8)}$ curves are due to the 
differential reddening affecting the cluster. Then, relative to V1, $\Delta
E(V-I)^{\phi(0.5-0.8)} = (V-i)^{\phi(0.5-0.8)}_{var} - (V-i)^{\phi(0.5-0.8)}_{V1}$
which in turn is used to calculate $\Delta E(B-V)^{\phi(0.5-0.8)} = \Delta
E(V-I)^{\phi(0.5-0.8)} / 1.616$ for each RRab star. These colour excesses can be
compared with the independent values obtained from the reddening map $E(B-V)^{map}$.
In the bottom panel of Fig.~\ref{fig:red-VI} we show the comparison of $\Delta
E(B-V)^{\phi(0.5-0.8)}$ versus $\Delta E(B-V)^{map} = E(B-V)^{map}_{var} -
E(B-V)^{map}_{V1}$. The result
should be a linear relation with unit gradient. Clearly the differential reddening in
the Alonso-Garcia et al. maps and the estimated ones from the RR Lyrae stars $V-i$
curves in the 0.5-0.8 phase range are, within the uncertainties, consistent.

The final adopted values of $E(B-V)$ for all of the RR Lyrae stars are those
from the Alonso-Garc\'ia et al. (2012) reddening map and they are listed in column 12 of
Table \ref{tab:fourier_coeffs}.

\section{Variable stars in NGC~6333}

The globular cluster NGC~6333 has not been explored by many investigators in search of
variable stars. The first variable star was discovered by Shapley (1916).
Thirty-five years passed until twelve more variable stars were discovered by Sawyer
(1951) as part of a photographic survey. Sawyer (1951) also measured the periods of
the thirteen known variables and labelled them V1-V13. The variable stars in NGC~6333
were not studied again until Christine Clement and collaborators started to work on
the cluster another thirty-five years later publishing four papers of interest. One of
the papers (Clement \& Shelton 1996) reports on a search for new variables using CCD
data which was successful in yielding 8 new detections (V14-V21) and extracting a
light curve for V11 for the first time. The other papers in the Clement series
(Clement, Ip \& Robert 1986; Clement \& Walker 1991; Clement \& Shelton 1999) present
various analyses of the known variables at their times of writing.

\begin{figure}
\includegraphics[width=8.0cm,height=8.0cm]{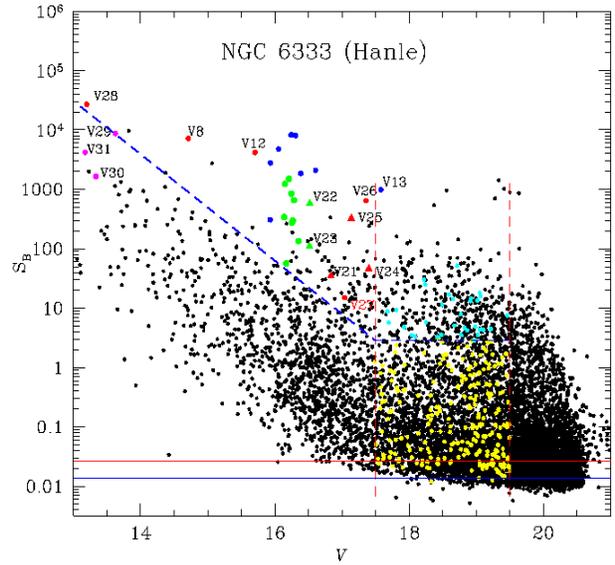}
\caption{Distribution of the $\cal S_{B}$ statistic as a function of mean $V$ magnitude
for 12519 stars measured in the $V$ Hanle images of NGC 6333. The coloured symbols
for variable stars are as described in the caption of Fig. \ref{CMD}. Stars in the Blue  
Straggler region with $\cal S_{B}$ below the variability detection threshold are plotted as yellow circles
while cyan circles represent Blue Straggler stars with $\cal S_{B}$ above the detection
threshold and hence potential variable candidates of the SX Phe type. However, none
of the Blue Stragglers was found to display convincing variability. The two vertical
dashed red lines correspond to
the magnitude limits set for the Blue Straggler region in the CMD.}
\label{alarm}
\end{figure}

\begin{figure} 
\includegraphics[width=8.0cm,height=8.0cm]{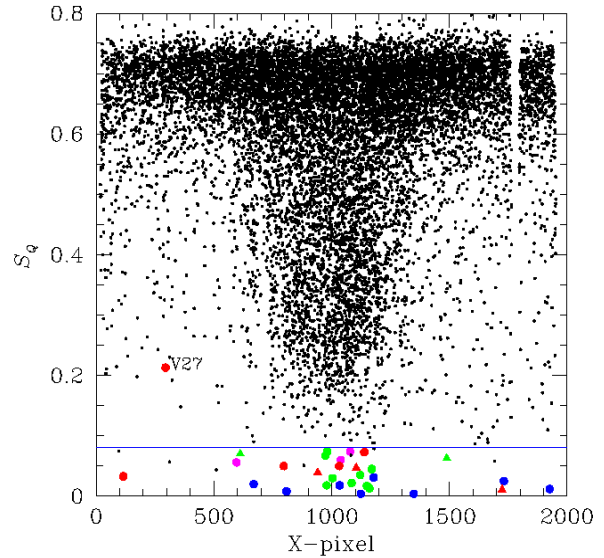}
\caption{Minimum value of the string-length parameter $S_Q$ calculated for the 12519
stars with a light curve in our $V$ reference image for the Hanle data, versus the CCD $x$-coordinate. The
coloured symbols are as described in the caption of Fig. \ref{CMD}.}
\label{SQ}
\end{figure}

In the following sections we describe the use of our time-series $Vi$ photometry to
search for new variables and to revisit the identifications, periodicities, and light curves
of the known variables.

\begin{table*}
\caption{General data for all of the confirmed variables in NGC~6333 in the FOV of 
our images. Stars V22-V34 are new discoveries in this work. Labels 'Bl' and 'd'
are for Blazhko variables and double mode RR Lyrae, respectively. The best previous
period estimates for each variable from the Clement series of papers (Clement, Ip \&
Robert 1984; Clement \& Walker 1991; Clement \& Shelton 1996; 1999) are reported in
column 10 for comparison with our refined periods in column 7.}
\label{variables}
\centering
\begin{tabular}{llllllllllll}
\hline
Variable & Variable & $<V>$   & $<i>$   & $A_V$       & $A_i$   & $P$ (days)    & $\beta$      & HJD$_{max}$     & $P$ (days)     & RA          & Dec.         \\
Star ID  & Type     & (mag)   & (mag)   & (mag)       & (mag)   & this work     & d Myr$^{-1}$ & (+2~450~000)    & Clement papers & (J2000.0)   & (J2000.0)    \\
\hline
V1       & RRab     & 16.276  & 16.520  & 1.155       & 0.758   & 0.5857309     &     
       & 5779.2366       & 0.585728       & 17:19:18.32 & --18:32:21.6 \\
V2       & RRab     & 16.209  & 16.391  & 1.110       & 0.683   & 0.6281843     &    
        & 5665.4329       & 0.628186       & 17:19:14.77 & --18:31:36.2 \\
V3       & RRab-Bl  & 16.364  & 16.586  & $>$1.00       &$>$0.70  & 0.605206      &  
 
        & 5664.4684       & 0.605353       & 17:19:26.41 & --18:34:34.8 \\
V4       & RRab     & 16.168  & 16.456  & 1.017       & 0.677   & 0.6713000     &     
       & 5723.4130       & 0.6713021      & 17:19:13.62 & --18:31:39.6 \\
V5       & RRc      & 16.256  & 16.548  & 0.449       & 0.285   & 0.3788136     &     
       & 5666.4490       & 0.378812       & 17:19:14.36 & --18:31:12.5 \\
V6       & RRab     & 16.351  & 16.478  & 1.000       & 0.672   & 0.6067809     &     
        & 5666.4452       & 0.607795       & 17:19:07.02 & --18:31:20.9 \\
V7       & RRab     & 16.566  & 16.655  & 1.120       & 0.715   & 0.6284626     &    
        & 5723.3838       & 0.6284615      & 17:19:04.18 & --18:32:24.4 \\
V8       & LPV      & 14.871  & 12.684  & $>$0.69     & $>$0.18 & -             &     
       & -               & 407            & 17:19:06.84 & --18:32:42.8 \\
V9       & RRc      & 16.261  & -       & $>$0.38     & -       & 0.3229883     &     
       & 6166.5212       & 0.322989       & 17:19:35.41 & --18:34:17.1 \\
V10      & RRc-Bl   & 16.285  & 16.674  & 0.449       & 0.313   & 0.3198454     &     
       & 6107.2557       & 0.319820       & 17:19:14.53 & --18:30:39.7 \\
V11      & RRab     & 16.065  & 16.300  & 0.704       & 0.543   & 0.7424499     &     
        & 5723.4130       & 0.73630        & 17:19:11.73 & --18:31:14.8 \\
V12      & An.Cep   & 15.671  & 15.765  & 0.92       & $>$0.51 & 1.340255      &     
       & 5323.3075       & 1.340204       & 17:18:52.69 & --18:33:20.3 \\
V13      & RRab     & 17.681  & 17.963  & 1.183       & 0.780   & 0.4798682     &     
       & 5323.3927       & 0.479874       & 17:19:30.29 & --18:30:54.6 \\
V14      & RRc      & 16.253  & 16.580  & 0.443       & 0.250   & 0.3270530     & 4.67
       & 5323.3708       & 0.32691        & 17:19:14.14 & --18:31:20.3 \\
V15      & RRab     & 16.239  & 16.343  & 0.971       & 0.760   & 0.6417673     &    
        & 5723.4130       & 0.64177        & 17:19:10.52 & --18:29:59.2 \\
V16      & RRc-Bl?  & 16.073  & 16.569  & 0.336       & 0.218   & 0.3846714    
&&5724.3724       & 0.38551        & 17:19:13.52 & --18:30:43.0 \\
V17      & RRc      & 16.243  & 16.593  & 0.375       & 0.283   & 0.3175888     &     
        & 5666.4294     & 0.31759        & 17:19:10.48 & --18:31:20.4 \\
V18      & RRc-Bl?  & 16.187  & 16.635  & 0.39       & 0.28    & 0.3413440     &11.50
       & 5323.2745       & 0.34228        & 17:19:10.61 & --18:30:41.9 \\
V19      & RRd:     & 16.264  & 16.580  & 0.47       & 0.29   & 0.3667937     &    
        & 5723.4130       & 0.36648        & 17:19:12.76 & --18:30:38.2 \\
V20      & RRc      & 16.340  & 16.591  & 0.423       & 0.219   & 0.3141782     &    
        & 5323.2668       & 0.31473        & 17:19:11.09 & --18:31:02.7 \\
V21      & EW       & 16.818  & 16.884  & 0.30$^{b}$ & 0.23$^{b}$   & 0.7204518    
&              & 5724.3785       & 0.360225       & 17:19:09.85 & --18:32:34.5 \\
V22      & RRc-Bl   & 16.483  & 16.740  & 0.510       & 0.261   & 0.3507561     &     
       & 5779.1641       & -              & 17:19:03.12 & --18:35:31.1 \\
V23      & RRc-Bl   & 16.513  & 16.827  & 0.311       & 0.178   & 0.3046535   &
            & 6108.1256       & -              & 17:19:21.15 & --18:30:10.2 \\
V24      & EW       & 17.381  & 17.520  & 0.45$^{b}$  & 0.380$^{b}$   & 0.366784     
&              & 6107.2525       & -              & 17:19:13.04 & --18:27:39.4 \\
V25      & E$^{c}$  & 17.111  & 17.333  & 1.642$^{b}$ &1.330$^{b}$ & -
    &              & 6107.2384$^{a}$ & -              & 17:19:25.94 & --18:27:43.9 \\
V26      & LPV      & 17.380  & 17.066  & $>$0.3      & $>$0.18 & -             &    
        & -               & -              & 17:19:13.84 & --18:29:37.4 \\
V27      & LPV      & 17.069  & 16.625  & $>$0.15     & $>$0.2  & -             &    
        & -               & -              & 17:18:56.40 & --18:32:47.4 \\
V28      & LPV      & 13.216  & 12.413  & $>$0.2      & $>$0.15 & -             &     
       & -               & -              & 17:19:11.68 & --18:31:04.5 \\
V29      & LPV      & 13.591  & -       & $>$0.2      & -       & -             &    
        & -               & -              & 17:19:02.68 & --18:32:53.6 \\
V30      & LPV      & 13.342  & 12.670  & $>$0.1      & $>$0.1  & -             &    
        & -               & -              & 17:19:11.83 & --18:31:27.9 \\
V31      & LPV      & 13.190  & 12.413  & $>$0.15     & $>$0.1  & -             &    
        & -               & -              & 17:19:12.65 & --18:31:01.7 \\
V32      & EW       & 15.961  & -       & 0.19        & -       & 0.17230$^{d}$ &     
      & 6162.6608       & -              & 17:19:37.10 & --18:35:40.1 \\
V33      & RRab     & 17.920  & -       & 0.517       & -       & 0.57597     &       
      & 6164.5019       & -              & 17:18:46.97 & --18:25:29.5 \\
V34      & LPV      & 15.045  & -       & $>$0.1      & -       & -             &     
        & -               & -              & 17:18:45.62 & --18:28:52.0 \\
\hline
\end{tabular}
\raggedright
\center{\quad $^{a}$ Time of minimum light. $^{b}$ Depth of eclipse. $^{c}$ Possibly EA. $^{d}$ Real period may be twice this value.}
\end{table*}

\subsection{Search for variable stars}
\label{sec:IDVAR}

We have been guided in the identification of the known variables by the finding charts
of Clement, Ip \& Robert (1984) and Clement \& Shelton (1996)
and by the equatorial coordinates of the variables in NGC~6333 given by Samus et al.
(2009). We had no problems identifying all 21 known variables in our own time-series data.
Then we applied a few approaches in the search for new variable star discoveries as we describe
below.

Firstly, we have defined a variability statistic $\cal S_{B}$ as:
\begin{equation}
{\cal
S_{B}}=\frac{1}{NM}\sum_{i=1}^{M}\left(\frac{r_{i,1}}{\sigma_{i,1}}+\frac{r_{i,2} }{
\sigma_{i,2}}+...+\frac{r_{i,k_{i}}}{\sigma_{i,k_{i}}}\right)^2,
\label{eq:sb}
\end{equation}
where $N$ is the total number of data points in the light curve and $M$
is the number of groups of time-consecutive residuals of the same
sign from the inverse-variance weighted mean magnitude. The residuals $r_{i,1}$ to
$r_{i,k_i}$ form the $i$th group of
$k_i$ time-consecutive residuals of the same sign with corresponding
uncertainties $\sigma_{i,1}$ to $ \sigma_{i,k_i}$. Fig \ref{alarm} shows the
distribution of the $\cal S_{B}$ statistic as a function of mean magnitude for the
12519 light curves for the stars in the Hanle $V$ images.

This statistic, based on the original
``alarm'' statistic $\cal A$ defined by Tamuz et al. (2006), has been used by
Arellano Ferro et al. (2012) to detect amplitude and period modulations in Blazhko variables.
Its application to detecting light curve variability was first introduced into our work by
Figuera Jaimes et al. (2013) where we discuss in detail its application and how to set theoretical
detection thresholds using simulated light curves.

As in Figuera Jaimes et al. (2013), we generated $10^5$ simulated light curves for
each star by randomly modifying the mean $\overline V$ magnitude
within the uncertainty $\sigma_i$ of each data point, i.e. $m_i=\overline V + {\lambda}_i {\sigma}_i$
where ${\lambda}_i$ is a random deviate drawn from a normal distribution with zero
mean and unit $\sigma$.
Then we used the resulting distribution of $\cal S_{B}$ values to determine the 50\% 
and 99.9\% percentiles which we plot in Fig.~\ref{alarm} as the horizontal blue and
red lines, respectively. Clearly the noise in the real Hanle $V$ light curves is not
Gaussian since many more than the expected $\sim$13 stars lie above the 99.9\%
percentile for pure Gaussian noise. Furthermore, for stars brighter than
$\sim$17.5~mag, the $\cal S_{B}$ statistic increases with brightness in an exponential
manner (linear in a logarithmic plot). All of these effects are due to residual
systematic errors in the light curves that can mimic real variability. It is clear
that our method of using simulated light curves to define the variability detection
threshold has not worked very well for our Hanle $V$ light curves of NGC~6333,
contrary to what we found for the NGC~7492 light curves from Figuera Jaimes et al.
(2013). This is because the $\cal S_{B}$ statistic is especially
sensitive to the systematic trends in the light curves, and these trends are simply
stronger and more coherent in the NGC~6333 light curves compared to the NGC~7492 light
curves.

We therefore opted to define our variability detection threshold by eye as the dashed blue line
in Fig.~\ref{alarm}. All known variable stars in the FOV of our Hanle images lie above our
chosen detection threshold by design. It is clear that the RR Lyrae stars have substantially larger values of
$\cal S_{B}$ among stars of their magnitude range. Long-term variables also stand clearly above the line.
We explored the light curves of all of the other stars above the threshold and could
identify seven clear new variables to which we assigned variable numbers; two RRc stars (V22 and V23), two
eclipsing binaries (V24 and V25), and three long period variables (V26, V27 and V28). Their classifications
and interesting properties will be discussed in $\S$ \ref{sec:IND_STARS}. Candidate variables in
the Blue Straggler region (see caption of Fig. \ref{alarm}) were investigated
individually but none showed convincing indications of variability. A similar plot as
in Fig. \ref{alarm} was constructed for our $V$ data from La Silla with very similar
results.

As a second strategy we also applied the string-length method (Burke, Rolland \&
Boy 1970; Dworetsky 1983) to each light curve to determine the period and
a normalized string-length statistic $S_Q$. In Fig. \ref{SQ} we plot the minimum $S_Q$
value for each light curve as a function of their corresponding CCD $x$-coordinate.
The known variables are plotted with the coloured symbols as described in the caption.
The horizontal blue line is not a statistically defined threshold but
rather an upper limit, set by eye, that contains the majority of the known variables.
Below this line we might expect to find previously undetected variables. In addition
to the known variables, there are ten other stars with $S_Q$ values below the
blue line and their light curves were thoroughly examined for variability.
We found long term variability for three of them, which we assign variable star
names
as V29, V30 and V31. We note that this method did not work for V27 but its variability
was confirmed by the analysis described below.

Finally, a third approach we have followed to identify variables in the field of our  
images is by detecting PSF-like peaks in a stacked image built from the sum of the 
absolute valued difference images normalized by the standard deviation in each pixel
as described by Bramich et al. (2011). This method allowed us to confirm the variability
of all of the new variables discovered so far and to find the new
variables V32, V33 and V34.

The previously known variables, along with all of the new discoveries, are listed in
Table~\ref{variables}.

The combination of the three approaches described above lead us to believe that our
search for variable stars with continuous variations (i.e. {\it not} eclipsing
binaries etc.) is fairly complete down to $V \sim 19$ for amplitudes
larger than 0.05 mag and periods between about 0.02d and a few hundred days.

\begin{figure*} 
\includegraphics[width=18.cm,height=14.cm]{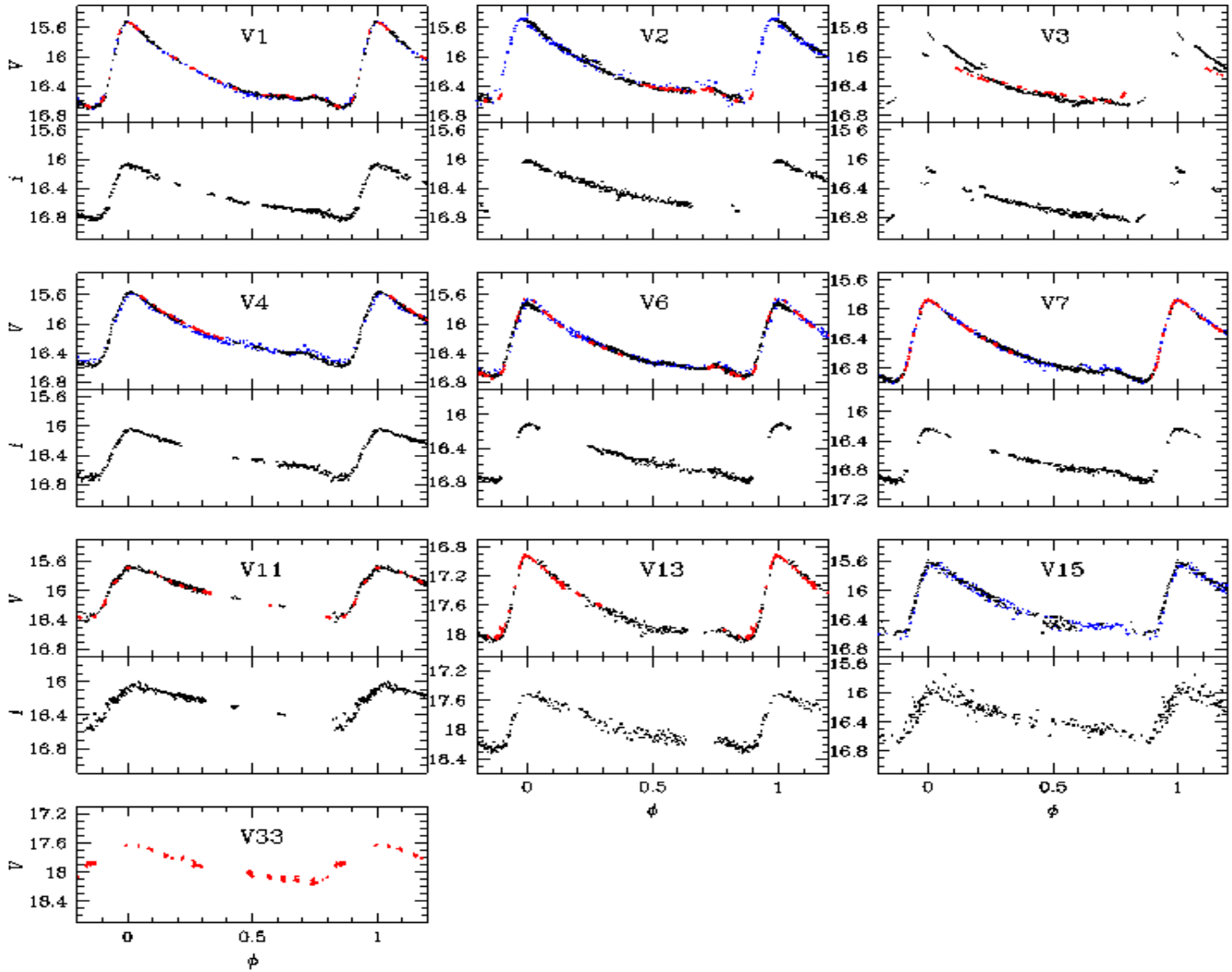}
\caption{Standard $V$ and instrumental $i$ light curves of the RRab stars in NGC~6333
phased with the periods listed in Table~\ref{variables}. Blue points represent $V$
data from Clement \& Shelton (1999). Black and red points represent Hanle and La Silla
data, respectively.}
    \label{VARSab}
\end{figure*}

\begin{figure*} 
\includegraphics[width=18.cm,height=15.cm]{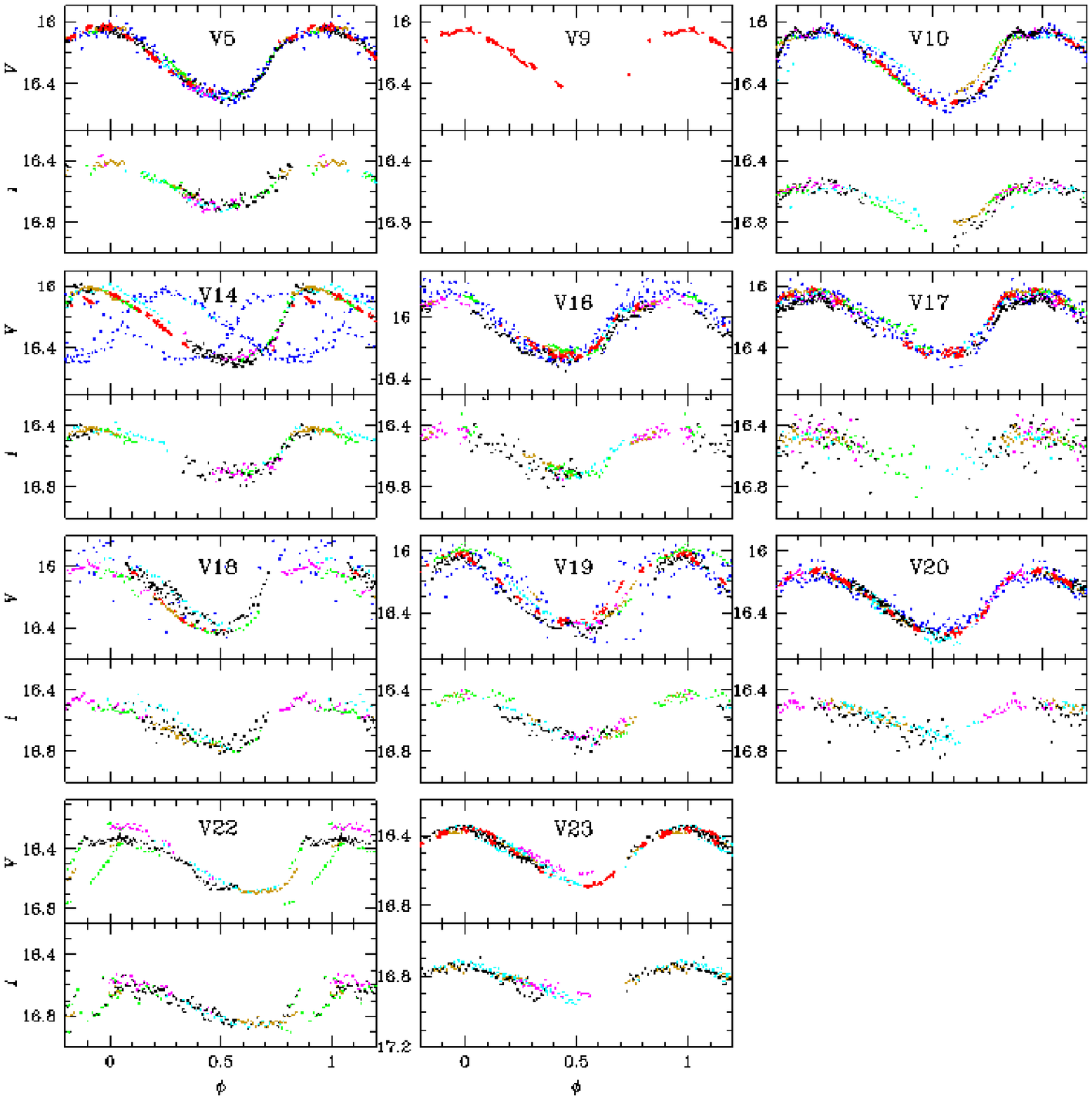}
\caption{Standard $V$ and instrumental $i$ light curves of the RRc stars in NGC~6333
phased with the periods listed in Table~\ref{variables}. To highlight any phase and
amplitude modulations, colours have been assigned for different observing runs:
May 1994 and 1995 - blue (Clement \& Shelton 1999), May 2010 - cyan, April 2011 - olive,
June-August 2011 - green, May 2012 - purple, June 2012 - black, August 2012 -
red.
The variable V9 is outside of the FOV of the Hanle images and thus only its $V$ light curve
from the La Silla data is displayed. Note that the light curves for V14 and V18 have been phased
with the best period determined without modelling a secular period change.}
    \label{VARSc}
\end{figure*}

\subsection{Period determination and refinement}
\label{sec:PERIOD}

To aid in the refinement of the periods of the known and newly discovered variables,
we have combined our data with the $V$ light curve data from Clement \& Shelton (1999).
We have noticed small zero point differences, of the order of a few hundreds of a magnitude,
between the three sets of data which may be different from star to star. This is to be
expected since, at least for DIA, the error in the reference flux affects all photometric
measurements for a single star from the same data set in the same way, and the data
for Hanle and La Silla each have a separate reference image with independently measured
reference fluxes. Thus for the period calculation we have proceeded as follows. First the
string-length method (Burke et al. 1970; Dworetsky 1983) was used to get a first estimate
of the period. Then small magnitude shifts were applied as necessary to the light curve data
for each star so as to better align the data and a second string length was run on the
levelled up light curve. The new period was used to phase the light curve and we
explored for further magnitude shifts if any. Generally two to three iterations were
sufficient to find an accurate period that phases the data precisely. 

The new periods and those from the Clement series of papers are given in columns 7 and 10
respectively of Table~\ref{variables}. We note that in general the agreement is good but the new
periods are considerably more precise. The light curves phased with the
new periods are displayed in Figs. \ref{VARSab} and \ref{VARSc}.

\subsection{Individual stars}
\label{sec:IND_STARS}

In this section we discuss the nature of some interesting known variables and the
newly discovered variable stars. To discuss their nature and cluster
membership we have built the colour magnitude diagram (CMD) of Fig.~\ref{CMD} by
calculating the inverse-variance weighted mean magnitudes of $\sim$11800 stars that have
both standard $V$ and instrumental $i$ magnitudes.

{\bf V3.} This is a RRab star for which a strong evidence of exhibiting the Blazhko effect
has been detected for the first time. Unfortunately our data do not fully cover the phased
light curve at different amplitudes (see Fig.~\ref{VARSab}).

{\bf V5.} Clement \& Walker 1991 report this star as undergoing a period change. However,
we do not detect this in our light curves combined with that from Clement \& Shelton (1999),
which have a combined baseline of $\sim$18 years (see Fig.~\ref{VARSc}).

\begin{figure*}
\includegraphics[width=18.cm,height=11.cm]{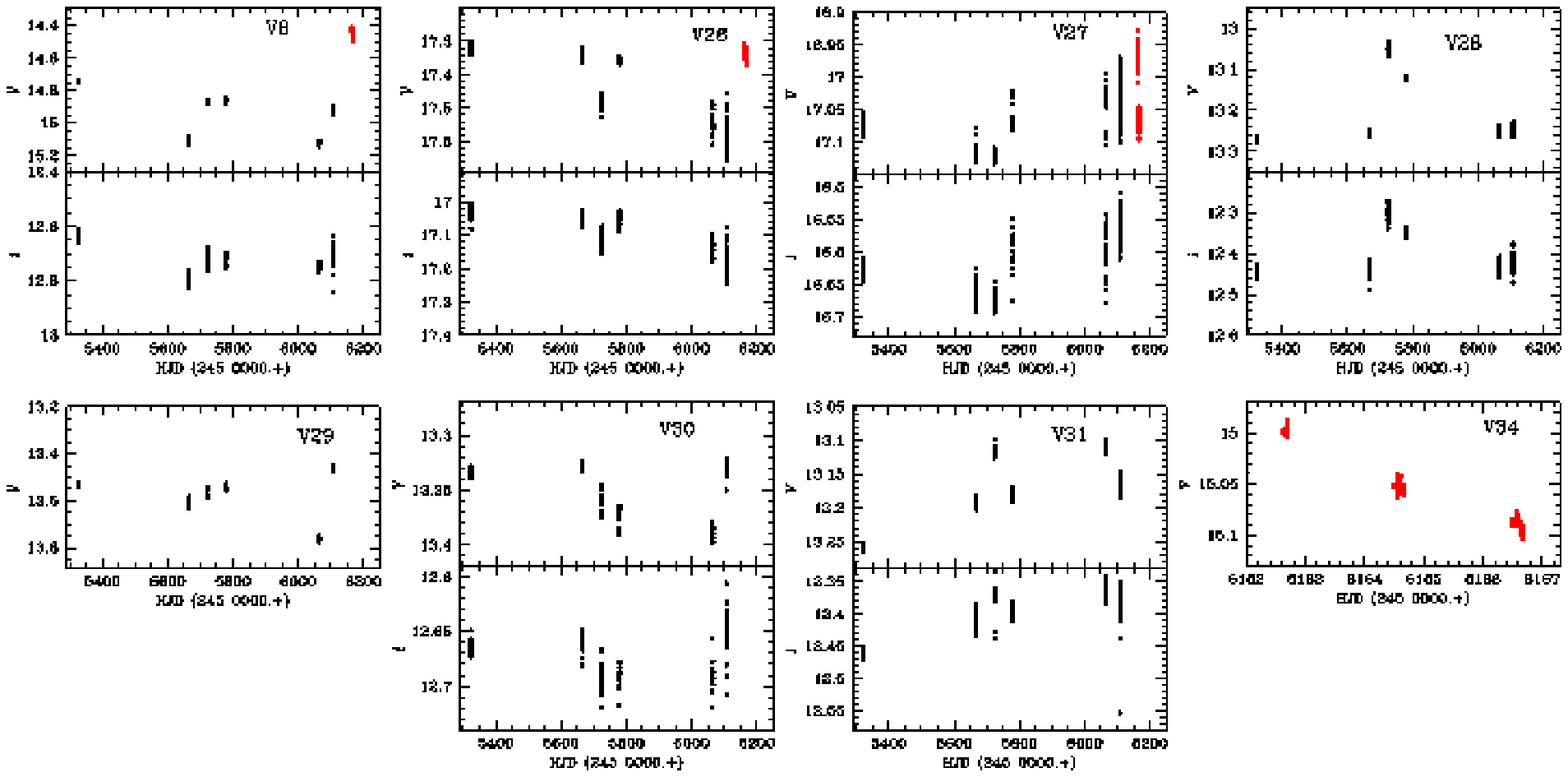}
\caption{$V$ and $i$ magnitude variations of the LPVs in NGC~6333. Black and red symbols are data from Hanle and La Silla,
respectively. V29 is saturated in our Hanle $I$-band images. V28, V29, V30 and V31 are saturated in our La Silla $V$-band
images. V34 is outside the FOV of our Hanle images.}
\label{LPV}
\end{figure*}

{\bf V8.} This variable was announced by Sawyer (1951) as a long period variable. A
period of ~407~d was suggested by Clement, Ip \& Robert (1984) and they commented on
its small $B$ amplitude of less that one magnitude. We show our light curves in $V$
and $i$ for this variable in Fig. \ref{LPV}. We are unable to estimate a period
although the two observed minima are indeed separated by about 400 days. More data are
required to complete the period analysis.
The position of the star in the CMD is much to the red of the RGB, even in the CMD corrected
for differential reddening, which implies that this
star is not a cluster member. We note that the reddening required to bring this star back to
the RGB if it is a cluster member is too large to be feasible for this field 
($E(B-V) \approx 1.5$~mag).

{\bf V10.} Clement \& Walker 1991 also report this star as undergoing a period change as a period increase.
Our light curve data show small phase and amplitude variations reminiscent of the Blazhko effect in RRc stars
(Arellano Ferro et al. 2012). Therefore we believe that it is the Blazhko effect that has been detected
previously and for which we present the first conclusive evidence of its presence in this star.

{\bf V12.} The variability of this star was discovered by Sawyer (1951) who finds it 
of similar brightness to the RR Lyrae stars. Clement, Ip \& Robert (1984) classified
it as a Population II cepheid with a period of 1.340204~d. They also find the star to
be of similar $B$ mag as the RR Lyrae stars which they use to
argue that the star is either much obscured by the presence of a prominent cloud to
the southwest of the cluster or that it is not a cluster member. We confirm its periodicity as
1.340255~d and we note that its position in the corrected CMD is about 1.05 magnitude
brighter than the mean HB, suggesting that indeed it is likely to be a cluster
member that suffers higher
than usual extinction due to the obscuring cloud to the SW of the cluster (see
Fig.~\ref{chart}). 
Clement et al. (2001) have pointed out that cepheids tend to occur in
globular clusters with blue horizontal branches. Assuming that V12 is indeed a cluster
member, and given the distance to the cluster (see $\S$ \ref{sec:distance}), its
absolute magnitude $M_V$ is $\sim -0.63$ mag which along with its period,
log~$P$=0.127, places the star on the P-L relation for anomalous cepheids (ACs)
pulsating in the fundamental mode (Pritzl et al. 2002, see their Fig. 6). ACs are
more luminous than Pop II cepheids for a given period,
they have a similar colour to RRc stars but are 0.5-1.5 mag brighter, and their period
can be between 0.5 and 3 days (Wallerstein \& Cox 1984). V12 fulfills all of these
characteristics, hence in the remainder of this paper we shall refer to V12 as an
AC.

{\bf V13.} This star was noted by Clement, Ip \& Robert (1984) to be much fainter than
the other RR Lyrae stars in the cluster and they consider it to be a field star. This is a clear
RRab star (see Fig.~\ref{VARSab}) and indeed it is $\sim$1.4~mag fainter than the other RRab stars.
While NGC 6333 is known for having heavy differential reddening, we discard interstellar extinction
as the cause of its faintness because its colour is similar to that of other RR Lyrae stars
(Fig. \ref {CMD}). Hence we agree with Clement, Ip \& Robert (1984) in arguing that
V13 is not a member of the cluster but that instead it is a background object.

{\bf V14.} The two periods 0.32691~d from Clement \& Shelton (1999) and
0.3270659~d from the string length method in this work fail to phase properly
the combined light
curve data from Clement \& Shelton (1999) and our light curves (Fig.\ref{VARSc}).
However, the
period 0.3270530~d accompanied with a secular period change rate of 4.67~d~Myr$^{-1}$ phases the light
curve much better as it will be shown in $\S$ \ref{Pdot}.

{\bf V16.} Clement \& Shelton (1996) speculate that this star is a double-mode RR Lyrae
(RRd). However, we have been able to phase the three available sets of $V$ data with one
single period 0.3846714~d, and we do not find any signs of secondary frequencies in the frequency
spectrum. Thus we do not confirm the double-mode nature of this RRc star (see
discussion in $\S$ \ref{2Mode}) but the star may be a Blazhko variable (see
Fig.~\ref{VARSc}).

{\bf V17.} The light curve of this star can be seen in Fig.~\ref{VARSc}. We note that
the celestial coordinates given by Samus (2009) seem to point
to the bright star near V17 while the authentic V17 is the more northern fainter star of the pair.
The correct coordinates are given in Table \ref{variables} and a proper
identification is in Fig.~\ref{chart}.

{\bf V18} The light curve of this RRc star displays nightly phase modulations which can be 
partially explained by a secular period change (see $\S$ \ref{Pdot}). However some modulations 
remain suggesting the presence of the Blazhko effect. Unfortunately, its light curve 
is noisy due to the position of the star in a heavily crowded region.

{\bf V19} The light curve of this RR Lyrae star displays nightly phase modulations 
since its light curve is not cleanly phased with the period found by the string-length method (see
Table \ref{variables}). It also seems to show some very mild amplitude variations (Fig. \ref{VARSc}).
The light curve is also reminiscent of the RRc stars with Blazhko effect found in M53 
by Arellano Ferro et al. (2012). In $\S$ \ref{2Mode} we shall discuss the possible
double mode nature of V19.

\begin{figure*} 
\includegraphics[width=15.0cm,height=6.0cm]{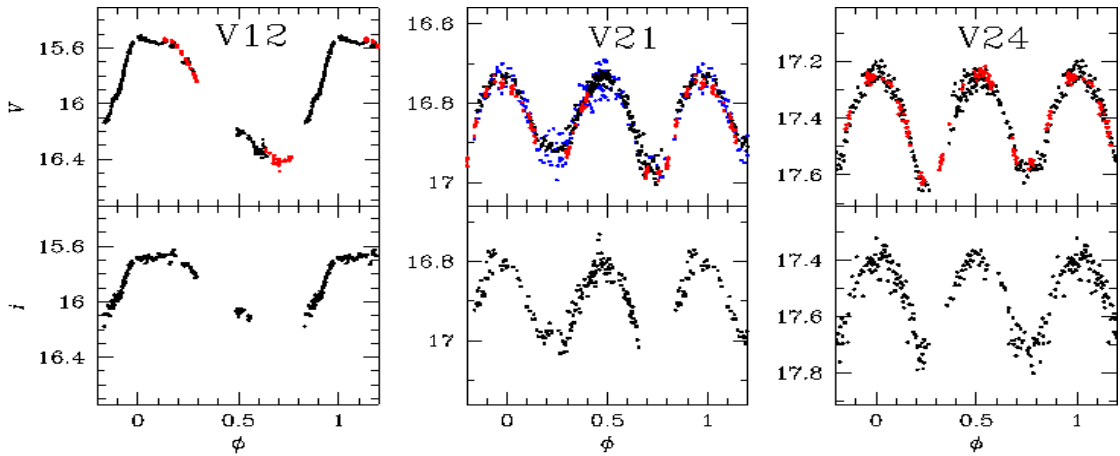}
\caption{$V$ and $i$ light curves of the anomalous cepheid V12, and the eclipsing
binaries
V21 and V24, phased with the periods given in Table~\ref{variables}. Black and red points
are data from Hanle and La Silla, respectively. The blue points correspond to the $V$ observations
of Clement \& Shelton (1999).}
\label{3STARS}
\end{figure*}

{\bf V21.} The variability of this star was discovered by Clement \& Shelton (1996).
These authors noticed that the star is fainter than the other RR Lyrae stars in the
cluster and that it has a substantially different Fourier $\phi_{21}$ parameter. Hence
they concluded that it is either not a RR Lyrae star or that it is not a cluster
member.
Later, Clement in the CVSGC classify it as an EW variable. In the corrected CMD the
star falls just below the HB by
$\sim$0.3 mag. Our best period is 0.7204518~d and it produces the light curve of Fig.
\ref{3STARS} where we notice two minima of different depths, typical of
semi-contact binaries. Thus we agree that the star is an EW star but with a period
approximately double the one reported in Clement \& Shelton (1996; 1999).
The finding chart in Clement \& Shelton 1996 is a bit misleading since the star is hardly visible 
in their map. This has probably led Samus (2009) to providing the wrong RA and Dec which
correspond to a neighbouring star. The correct coordinates are given in Table
\ref{variables}. The star is properly identified in our finding chart of
Fig. \ref{chart}.

{\bf V22 and V23.} We have discovered these two new variables. Their period, light
curves (see Fig.~\ref{VARSc}) and position
in the CMD lead us to classify them as RRc stars. The light curve of V22 shows strong 
phase and amplitude modulations, while the light curve of V23 also shows phase and
amplitude modulations to a lower level.
We therefore conclude that both stars may exhibit the Blazhko effect. However in
$\S$ \ref{2Mode} we shall explore the double mode possibility by searching for a
secondary frequency in the spectrum.

{\bf V24.} This is a new eclipsing binary with a period of 0.366784~d with two
different
eclipse depths (see Fig.~\ref{3STARS}). Its period along with its position in the corrected CMD
$\sim$1.2 mag below the HB support its classification as a W Ursae Majoris-type binary (or EW).

\begin{figure} 
\includegraphics[width=7.5cm,height=7.5cm]{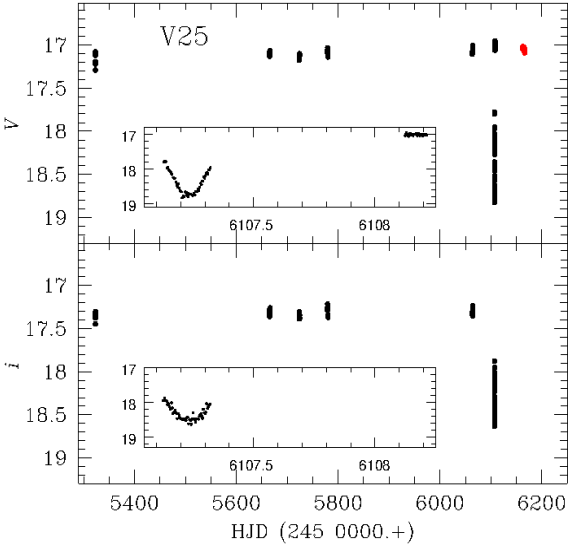}
\caption{Light curve of V25 where one eclipse has been detected at HJD 2456107.24d.
The
inset panel is a zoom-in on the eclipse. Black and red points represent Hanle and La
Silla
data respectively.}
\label{fig:V25HJD}
\end{figure}

{\bf V25.} The light curve of this new eclipsing binary is shown in Fig. \ref{fig:V25HJD}.
One eclipse has been detected at HJD~2456107.24 in both the $V$ and $I$ filters. With only
one eclipse we are unable to estimate the orbital period. We note that there is also the
hint of ellipsoidal variations in the out-of-eclipse light curve.

{\bf V26, V27, V28, V29, V30, V31 and V34.} The light curves of these new variable stars show long-term
variations (Fig~\ref{LPV}). For V29 and V34 we have no I-band data. However, the rest
of these variables are located well within the RGB in the CMD (Fig. \ref{CMD}). Our data are not sufficient
to calculate their periods.

\begin{figure} 
\includegraphics[width=7.0cm,height=7.5cm]{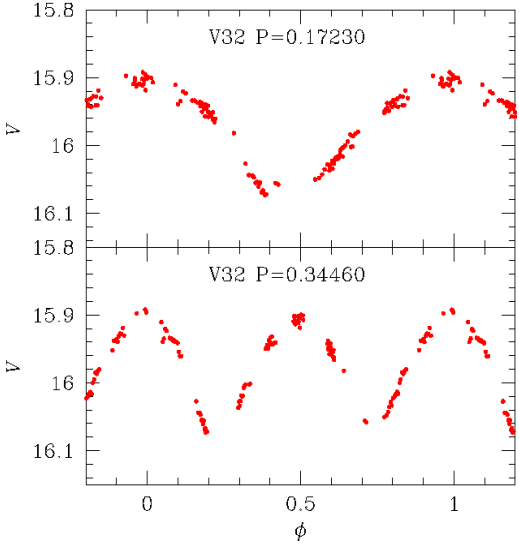}
\caption{Light curve of the binary star V32 phased with two possible periods. The 
star is out the FOV of Hanle, hence we can only plot data from La Silla.}
\label{fig:V32}
\end{figure}

{\bf V32} The light curve of this new variable is shown in Fig.\ref{fig:V32} phased with
our best period found by the string-length method of 0.17230~d. Twice this period
would produce a clean double-wave light curve with some suggestion of different depth
minima. The star is
outside the FOV of our Hanle images and therefore we only have $V$-band data from La
Silla.
Although its mean magnitude $V \sim$16 mag is similar to the brightness of the RRc
stars in the cluster,
we believe the star is rather an eclipsing binary of the W Ursae Majoris-type (or
EW), whose period
and nature can be better defined upon obtaining further accurate data. 

{\bf V33} The light curve phased with our best period 0.57540~d is shown in
Fig.~\ref{VARSab}. This new variable star is obviously an RRab star but given its mean magnitude
$V \sim$17.94 mag it must be a field star further away than NGC 6333. The star is
outside the FOV of our Hanle images.

{\bf V34} This new variable is a long-term pulsator whose variability is clear in the three nights
of data as shown in Fig.~\ref{LPV}. The star is outside the FOV of our Hanle
images.

\subsection{Double mode pulsators}
\label{2Mode}

Clear phase and/or amplitude modulations are seen in several of the RRc light curves
in Fig.~\ref{VARSc}. Phase and amplitude modulations can be the result of a
Blazhko effect or a double mode pulsation. To distinguish between these possibilities
one requires a convincing identification of secondary periods
in the light curve and a long time-series of accurate photometry is generally
needed. Despite the limitations of our data set we have attempted the
identification of such secondary periods for the RRc stars showing, to some extent,
phase modulations, i.e. V10, V16, V18, V19, V22 and V23.

\begin{figure} 
\includegraphics[width=8.5cm,height=7.0cm]{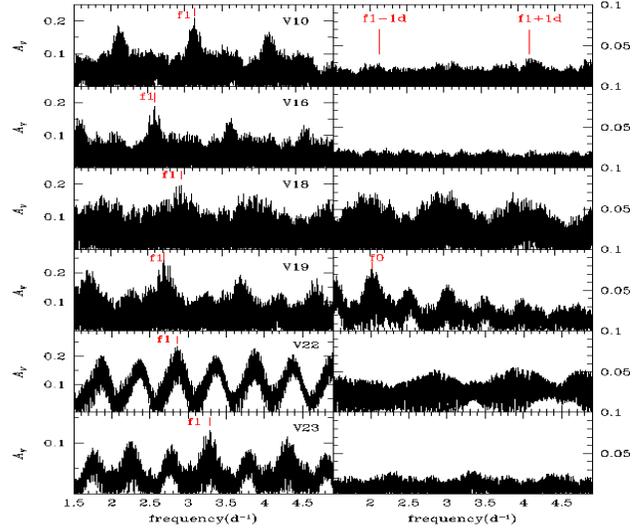}
\caption{Frequency spectra of selected RRc stars. The left-hand panels show the
spectrum
produced by the original data. The major peaks correspond to the periods listed in
Table~\ref{variables} and used to produce the light curves in Figs. \ref{VARSab} and 
\ref{VARSc}.
The right-hand panels show the spectra of the residuals after pre-whitening the main
frequency.
Note that the vertical scale in the right-hand panels has been increased to highlight
possible
secondary frequencies. See text for a discussion.}
\label{fig:DOUMODE}
\end{figure}

We used the program {\tt period04} (Lenz \& Breger 2005) to identify the primary or
first overtone period $P_1$ 
previously found by the string-length method described in section \ref{sec:PERIOD} and
given in column 7 of Table
\ref{variables}. Then we prewhitened the data from the primary period and searched
the residuals for a secondary period. 
The frequency spectra of the original light curves and the residuals are
shown in Fig. \ref{fig:DOUMODE}. No significant secondary frequencies were detected
in the residual spectra of V10, V16, V18, V22 and V23 other than small residuals at
the 1-day aliases of the main frequency f1. For V22 however, the amplitude modulations
are so prominent that they could not be explained by period variations and rather they
must be the result of the Blazhko effect whose periodicity we are not in
position to estimate given our data set.

\begin{figure} 
\includegraphics[width=7.5cm,height=4.0cm]{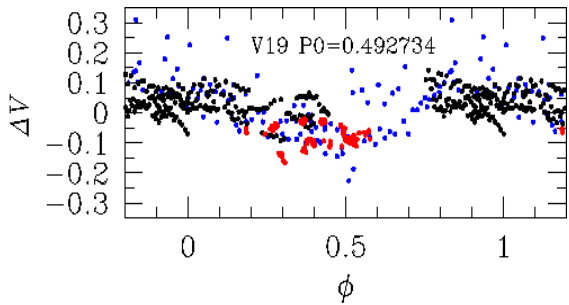}
\caption{Residuals of V19 phased with the fundamental period given in the legend. Blue
points are the $V$ data from Clement \& Shelton (1999). Black and red points are
the $V$ data from Hanle and La Silla, respectively. See $\S$ \ref{2Mode} for a
discussion.}
\label{DLC}
\end{figure}

For V19 a rather prominent frequency was found in the residual spectrum at
2.02949d$^{-1}$,
or a period of
0.492734d. If this period is interpreted as the
fundamental $P_0$ and with $P_1$=0.3667937d we find a ratio $P_1/P_0 =0.744$ which
corresponds to the canonical
0.746$\pm$0.001 ratio in RRd stars (Catelan 2009, Cox, Hudson \& Clancy 1983) and with
the period ratio found in a large sample of double mode RR Lyrae stars in the LMC
(Alcock et al. 2000). $P_0$ produces a residuals light curve shown in  Fig. \ref{DLC}.
The above two facts strongly suggest that V19 is indeed a double mode
RR Lyrae or RRd star. An inspection of the light curve
of V19 in Fig. 4b of Clement \& Shelton (1996) reveals clear nightly
phase drifts like those noted by these authors for V16, however this
case was not pursued further by them.

Thus, V10, V16, V18, V22 and V23 are rather reminiscent of the Blazhko RRc
variables in NGC~5024 (Arellano Ferro et al. 2012). Given the nature of our
time-series we cannot estimate their Blazhko period or the possible presence of
non-radial modes for which dense, accurate and
prolonged observations are required as recent experience has shown in targets of space
missions (e.g. Guggenberger el al. 2012 and references there in).

\subsection{RRc stars with secular period variation}
\label{Pdot}

The RRc stars V14 and V18 show the largest phase variations not obviously
accompanied with amplitude modulations. This is suggestive of a secular period
change. To investigate this possibility we have used a variation of the
string-length method previously described in Bramich et al. (2011). We define:

\begin{equation}
\label{eq:chanceperiod}
 \phi(t)=\frac{t-E}{P(t)}-{\left \lfloor \frac{t-E}{P(t)} \right \rfloor}
\end{equation}

\begin{equation}
 P(t)=P_0+\beta(t-E),
\end{equation}

\noindent
where $\phi$(t) is the phase at time $t$, $P(t)$ is the period at time $t$, $P_0$ is
the period at the epoch $E$, and $\beta$ is the rate of period change. We fix the
value of $E$ and calculate the best-fitting values of $P_0$ and $\beta$ (in units d
d$^{-1}$) within a small range of possible periods around the previously found
best-fitting period as described in $\S$ \ref{sec:PERIOD}. We have applied this
approach to the light curves of V14 and V18, both of which show clear
phase displacements over time (see $\S$ \ref{sec:IND_STARS}).

\begin{figure} 
\includegraphics[width=8.0cm,height=8.0cm]{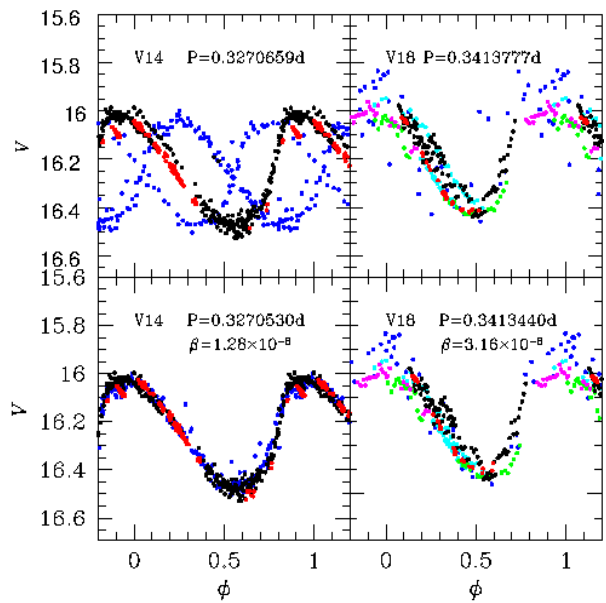}
\caption{Two RRc stars with secular period change. The top panels show the light
curves phased with a constant period found as described in $\S$\ref{sec:PERIOD}. The
bottom panels show the light curves phased with the new period and period change rate
$\beta$ given in the legend. The colours are coded as in the caption of
Fig.~\ref{VARSc}.}
\label{fig:lcpdot}
\end{figure}

In Fig. \ref{fig:lcpdot} we show the light curves of these two stars phased with a
constant period (top panels), and with the new period and period change
rate calculated with the above equations (bottom panels). It is clear that the new
periods and period change rates produce much cleaner and more coherent light curves. We
conclude that V14 and V18 have secular period changes at the rates of 4.67
and 11.5~d~Myr$^{-1}$, respectively. In the case of V18, the light curve phasing is
still not fully satisfactory and therefore we do not discard the possibility of additional amplitude
modulations that could be associated with a Blazhko effect similar to many of the RRc
stars in NGC 5024 (Arellano Ferro et al. 2012).

\begin{figure*}
 \centering
\epsfig{file=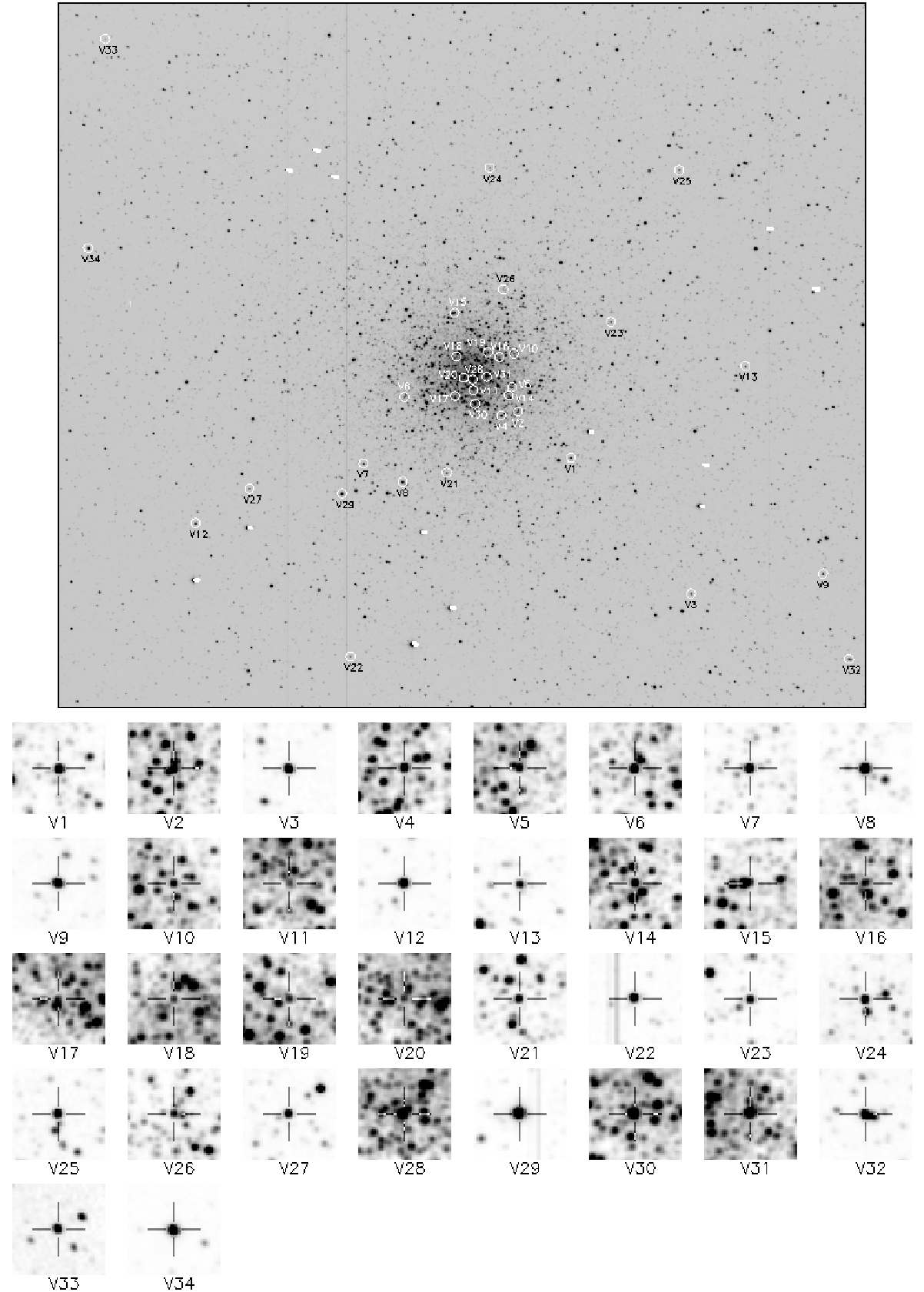,angle=0.0,width=0.95\linewidth}
\caption{Finding charts constructed from our La Silla $V$ reference image; north is up
and east is to the right. The cluster image is 13.04$\times$11.39~arcmin$^{2}$, and
the image stamps are of size 20.6$\times$20.6~arcsec$^{2}$. Each confirmed variable
lies at the centre of its corresponding image stamp and is marked by a cross-hair.}
\label{chart}
\end{figure*}

\subsection{Search for variable stars among the blue stragglers}

The blue straggler (BS) region in NGC~6333 is arbitrarily defined in Fig.~\ref{CMD} by the
dashed red lines. In Fig. \ref{alarm}, this translates to the magnitude limits indicated
by the two vertical red dashed lines. In this figure, all of the stars in the BS region
are plotted with coloured points; yellow or cyan depending on whether they lie below or
above the chosen variability detection threshold for $\cal S_{B}$. We found that
40 of the BS stars have the $\cal S_{B}$ statistic for the Hanle data above this
threshold and their light curves were explored in detail. However, no clear and convincing variability
was found in any of these stars and the relatively high value of $\cal S_{B}$ could always be explained
by groups of spurious photometric measurements or by the effect on the photometry of a
nearby known
variable. We conclude that if SX Phe stars exist in this cluster, then they must be of an amplitude
similar to or smaller than the RMS achieved by our photometry (see Fig. \ref{rms}).
To highlight this point, if we consider that the
known SX Phe stars in NGC~5024 (another OoII cluster, with a blue HB and similar
metallicity to NGC~6333)
(Arellano Ferro et al. 2011) actually existed instead in NGC~6333, then we would have 
been able to detect about 19 out of 25 (~76\%) of them in our data, given the
completeness of our variable star search reported at the end of $\S$ \ref{sec:IDVAR}.

\section{RR Lyrae stars}
\label{sec:RRLstars}

\subsection{[Fe/H] and M$_V$ from light curve Fourier decomposition}
\label{sec:FEMV}
 
Estimates of physical parameters, such as metallicity, luminosity
and effective temperatures can be made from the Fourier decomposition of
the light curves of RR Lyrae stars into their harmonics and from semi-empirical
relationships (e.g. Jurcsik \& Kov\'acs 1996, Morgan et al. 2007). Traditionally the
light curves are
represented by the equation:

\begin{equation}
m(t) = A_o ~+~ \sum_{k=1}^{N}{A_k ~\cos~( {2\pi \over P}~k~(t-E) ~+~ \phi_k ) },
\label{eq_foufit}
\end{equation}

\noindent
where $m(t)$ are magnitudes at time $t$, $P$ the period and $E$ the epoch. A linear
minimization routine is used to fit the data with the Fourier series model, deriving
the best fit values of the amplitudes $A_k$ and phases $\phi_k$ of the
sinusoidal components. 
From the amplitudes and phases of the harmonics in eq. \ref{eq_foufit}, the Fourier
parameters $\phi_{ij} = j\phi_{i} - i\phi_{j}$ and $R_{ij} = A_{i}/A_{j}$ are defined. 
Although the $V$ data from Clement \& Shelton (1999) have been very useful in
refining the periods of the RR Lyrae stars, in fitting the light curves we have opted
not to include the data since they have a considerably larger scatter. The mean
magnitudes $A_0$, and the Fourier light curve fitting parameters of the
individual RRab and RRc type stars in $V$ are listed in Table
~\ref{tab:fourier_coeffs}. 

\begin{table*}
\caption{Fourier coefficients $A_{k}$ for $k=0,1,2,3,4$, and phases $\phi_{21}$,
$\phi_{31}$ and $\phi_{41}$, for the 9 RRab and 7 RRc type variables for which 
the Fourier decomposition fit was successful. The numbers in parentheses indicate
the uncertainty on the last decimal place. Also listed are the number of
harmonics $N$ used to fit the light curve of each variable, the deviation 
parameter $D_{\mbox{\scriptsize m}}$ (see Section~\ref{sec:FEMV}) and the colour
excess $E(B-V)$.}
\centering                   
\begin{tabular}{lllllllllrrc}
\hline
Variable ID     & $A_{0}$    & $A_{1}$   & $A_{2}$   & $A_{3}$   & $A_{4}$   & $\phi_{21}$ & $\phi_{31}$ & $\phi_{41}$ 
& $N$   & $D_{\mbox{\scriptsize m}}$&$E(B-V)$ \\
     & ($V$ mag)  & ($V$ mag)  &  ($V$ mag) & ($V$ mag)& ($V$ mag) & & & & & &  (mag)\\
\hline
       &       &   &   &   & RRab stars    & &        
   &             &             &       &          \\
\hline
V1 & 16.276(1)& 0.401(1) &0.198(1) &0.141(1)& 0.095(1)&3.910(8)& 8.118(12)&
6.109(16)& 9  & 1.9&0.416\\
V2 & 16.209(1)& 0.372(2)&0.187(2) & 0.129(2) & 0.093(2) &3.863(14) &8.167(22)
&6.147(31) & 9  & 2.4&0.378\\
V4 & 16.168(1)&0.341(1)  &0.184(1)   &0.119(1)   &0.071(1)  &4.124(11)   
&8.495(19) &6.664(27) &9  &0.8&0.378\\
V6 & 16.351(1) &0.341(2)   &0.175(2)   &0.119(2)   &0.079(2)  &3.965(14)    
&8.194(21)  &6.121(30)  &9  &2.7&0.422\\
V7 & 16.566(1) &0.379(1)   &0.187(1)   &0.130(1)   &0.089(1)  &3.941(9)    
&8.209(12)  &6.226(17)  &9  &0.7&0.489\\
V11 & 16.065(1) &0.252(4)   &0.129(4)   &0.074(4)   &0.031(4)  &4.410(43)    
&8.903(67)  &7.210(137)  &7   &4.9&0.387 \\
V13 & 17.681(3) &0.403(4)   &0.186(4)   &0.139(4)   &0.097(3)  &3.804(31)    
&7.961(42)  &5.870(64)  &8  &3.3&0.443\\
V15 & 16.239(3) &0.350(4)   &0.186(4)   &0.139(4)   &0.074(4)  &3.810(32)    
&8.060(49)  &6.100(68)  &8  &3.7&0.411\\
V33 & 17.904(3) &0.212(5)   &0.090(4)   &0.025(4)   &0.016(5)  &4.407(75)    
&8.674(191)  &7.929(271)  &4  &3.1&0.507\\
\hline
             &            &           &           &           & RRc stars &           
 &             &             &       & &\\
\hline
V5  &16.256(1)  &0.225(1)   &0.028(1)   &0.012(1)   &0.008(1)  &4.828(47)    
&3.932(105)  &2.206(150)  &4& ---&0.388\\
V10 &16.285(2)  &0.232(3)   &0.033(3)   &0.013(3)   &0.017(3)  &5.013(87)    
&2.746(241) &1.390(164)  &4& ---&0.407\\
V14 &16.253(2)  &0.221(3)   &0.048(3)   &0.024(3)   &0.021(3)  &5.073(70)   
&3.241(126)  &1.523(157)  &4& ---&0.382\\
V16 &16.073(2)  &0.188(2)   &0.015(2)   &0.025(2)   &0.013(2)  &3.919(137)   
&4.353(103)  &1.580(157)  &4& ---&0.395\\
V17 &16.243(2)  &0.187(2)   &0.044(2)   &0.007(3)   &0.016(2)  &4.921(74)    
&3.894(400) &1.747(166)     &4& ---&0.403\\
V20 &16.340 (1)  &0.211(1)   &0.036(2)   &0.012(2)   &0.001(1)   &4.886(46)   
&2.967(123) &1.821(550)     &4& ---&0.396\\
V23 &16.513(1)  &0.149(2)   &0.024(2)   &0.007(2)   &0.004(2)   &4.891(73)   
&4.619(249) &2.684(294)     &4& ---&0.457\\
\hline
\end{tabular}
\label{tab:fourier_coeffs}
\end{table*}

The Fourier decomposition parameters can be used to calculate [Fe/H] and M$_V$ for
both RRab and RRc stars
by means of the semi-empirical calibrations given in eqs. \ref{eq:JK96}, \ref{eq:ḰW01},
\ref{eq:Morgan07} and \ref{eq:K98}. 

The calibrations for [Fe/H] and M$_V$ used for RRab stars are:

\begin{equation}
{\rm [Fe/H]}_{J} = -5.038 ~-~ 5.394~P ~+~ 1.345~\phi^{(s)}_{31},
\label{eq:JK96}
\end{equation}

\begin{equation}
M_V = ~-1.876~\log~P ~-1.158~A_1 ~+0.821~A_3 + K,
\label{eq:ḰW01}
\end{equation}

\noindent given by Jurcsik \& Kov\'acs (1996) and Kov\'acs \& Walker (2001),
respectively.  The standard deviations of the above calibrations are 0.14 dex (Jurcsik
1998) and 0.04 mag, respectively. In eq. \ref{eq:ḰW01} we have used K=0.41 to scale the
luminosities of the RRab with the distance modulus of 18.5~mag for the LMC (see the
discussion in Section~4.2 of Arellano Ferro et al. 2010).
Eq. \ref{eq:JK96} is applicable to RRab stars with a {\it deviation parameter} $D_m$,
defined by Jurcsik \& Kov\'acs (1996) and Kov\'acs \& Kanbur (1998), not exceeding an
upper limit. These authors suggest $D_m \leq 3.0$. The $D_m$ is listed in
column 11 of Table~\ref{tab:fourier_coeffs}. A few stars have $D_m$ marginally larger
than this limit but given the quality of their light curve and the good coverage of
the cycle we opted for reporting their iron abundance.
The metallicity scale of eq. \ref{eq:JK96} was transformed into the widely used scale
of Zinn \& West (1984)
using the relation [Fe/H]$_{J}$ = 1.431[Fe/H]$_{ZW}$ + 0.88 (Jurcsik 1995). These two
metallicity scales closely coincide for [Fe/H]$\sim -2.0$ while for [Fe/H]$\sim -1.5$,
the [Fe/H]$_{J}$ is about 0.24 dex less metal poor than [Fe/H]$_{ZW}$ (see also Fig. 2
of Jurcsik 1995). Therefore, for a metal poor cluster such as NGC 6333, the two scales
are not significantly different. 

\begin{table*}
\footnotesize
\begin{center}
\caption[] {\small Physical parameters for the RRab and RRc stars. The numbers in
parentheses indicate the uncertainty on the last 
decimal place and have been calculated as described in the text.}
\label{fisicos}
\hspace{0.01cm}
 \begin{tabular}{lcccccc}
\hline 
Star&[Fe/H]$_{ZW}$ & $M_V$ & log$(L/{\rm L_{\odot}})$ & log~$T_{\rm eff}$  &
$M/{\rm M_{\odot}}$&$R/{\rm R_{\odot}}$\\
\hline
 &  &  & RRab stars &  & & \\
\hline
V1 &$-1.666(11)$&0.497(1)&1.701(1)&3.808(7)& 0.73(6)&5.77(1) \\
V2 &$-1.780(21)$&0.464(3)&1.714(1)&3.803(8)& 0.73(7)&6.00(1)\\
V4 &$-1.634(18)$&0.438(1)&1.725(1)&3.800(8)& 0.70(6)&6.15(1)\\
V6 &$-1.674(20)$&0.520(3)&1.692(1)&3.805(8)& 0.70(6)&5.78(1)\\
V7 &$-1.742(11)$&0.456(1)&1.717(1)&3.803(7)& 0.73(6)&6.01(1)\\
V11 &$-1.519(63)$&0.422(6)&1.731(2)&3.794(17)& 0.66(13)&6.38(4)\\
V13$^a$ &$-1.415(43)$&0.656(6)&1.638(2)&3.819(10)&0.71(8)&5.09(2)\\
V15&$-1.932(46)$&0.456(6)&1.717(2)&3.798(10)& 0.75(9)&6.14(2)\\ 
V33$^a$&$-1.107(181$)&0.635(7)&1.646(3)&3.803(31)&0.68(25)&5.54(7)\\
\hline
Weighted &&&&&&\\
Mean & -1.702(6)& 0.467(1) & 1.713(1) &3.803(3) & 0.72(3)&5.73(1) \\
\hline 
 &  &  & RRc stars &  & & \\
\hline
V5  &$-1.81(22)$ & 0.518(2)& 1.693(1)& 3.885(1) &0.49(1) &4.54(7)\\
V10 &$-1.83(44)$ & 0.527(4)& 1.689(2)& 3.862(1)&0.59(1) &4.44(13)\\
V14 &$-1.69(24)$ & 0.499(3)& 1.700(1)& 3.863(1)&0.58(1) &4.46(7)\\
V16 &$-1.65(23)$ & 0.530(6)& 1.688(3)& 3.859(1)&0.46(1) &4.48(7)\\
V17 &$-1.22(78)$ & 0.537(3)& 1.685(1)& 3.869(2)&0.54(2) &4.28(20)\\
V20 &$-1.69(22)$ & 0.609(2)& 1.656(1)& 3.864(1)&0.54(1) &4.23(7)\\
V23 &$-0.45(50)^a$& 0.604(3)& 1.658(1)& 3.875(1)&0.49(1) &4.02(11)\\
\hline
Weighted&&&&&\\
Mean & -1.71(11) & 0.554(1)  & 1.638(1) & 3.862(1) &0.51(1) &4.79(4) \\
\hline
\hline
\end{tabular}
\end{center}
$^a$Values not included in the average. V13 and V33 are not cluster members.
\end{table*}

For the RRc stars we employ the calibrations: 

$$ {\rm [Fe/H]}_{ZW} = 52.466~P^2 ~-~ 30.075~P ~+~ 0.131~\phi^{(c)~2}_{31}  $$
\begin{equation}
~~~~~~~	~-~ 0.982 ~ \phi^{(c)}_{31} ~-~ 4.198~\phi^{(c)}_{31}~P ~+~ 2.424,
\label{eq:Morgan07}
\end{equation}

\begin{equation}
M_V = 1.061 ~-~ 0.961~P ~-~ 0.044~\phi^{(s)}_{21} ~-~ 4.447~A_4, 
\label{eq:K98}	
\end{equation}

\noindent given by  Morgan et al. (2007) and Kov\'acs (1998), respectively.
The standard deviations of the above calibrations are 0.14 dex and 0.042 mag
respectively. For eq. \ref{eq:K98} the zero point was reduced to 1.061 mag to make the
luminosities of the RRc consistent with the distance modulus of 18.5~mag for the LMC (see
discussions by Cacciari et al. 2005 and Arellano Ferro et al. 2010). The original zero
point given by Kov\'acs (1998) is 1.261.

In the above calibrations the phases are calculated either from series of sines or of
cosines as indicated by the superscript. We transformed our 
cosine series phases into the sine ones where necessary via the relation ~~
$\phi^{(s)}_{jk} = \phi^{(c)}_{jk} - (j - k) {\pi \over 2}$.

The physical parameters for the RR Lyrae stars are reported in Table~\ref{fisicos}. 
We have not included the star V3 since it has prominent Blazhko modulations and
V9 because our observations are not sufficient to cover the complete light curve. 
Despite the fact that most RRc stars show to some extent amplitude
and/or phase modulations, in calculating mean
parameters, we only excluded stars with extreme modulations; namely V18 and V22.
V19 was also not considered given its double-mode nature. We also excluded V13 and
V33 which are not cluster members. V14 was considered only after the light curve
was phased with the period change rate included, i.e. the light curve in the bottom
panel of Fig. \ref {fig:lcpdot}. The inverse variance square weighted means are also
given in Table~\ref{fisicos}. The systematic error in the metallicity estimates 
is of the order of the scatter in the calibrations of eqs. \ref{eq:JK96} and
\ref{eq:Morgan07}, i.e. 0.14 dex.
Thus, the metallicity obtained from the RRab and RRc stars is
[Fe/H]$_{ZW}=-1.70\pm0.01$ which can be converted to 
the new scale defined by Carretta et al. (2009) using UVES spectra of RGB stars in
globular clusters by [Fe/H]$_{UVES}$=
$-0.413$ +0.130[Fe/H]$_{ZW}-0.356$[Fe/H]$_{ZW}^2$. 
We find [Fe/H]$_{UVES}=-1.67 \pm 0.01$. Clement \& Shelton (1999) found, from the
light
curve Fourier decomposition of V2, V4, V6 and V7, the average 
[Fe/H]$_{ZW}=-1.77\pm0.08$ in good agreement with our result.

To the best of our knowledge no iron abundance of NGC~6333 has been calculated from
high resolution spectroscopy. The first calculation of [Fe/H]$_{ZW}=-1.81\pm0.15$
was made from integrated photometry in the Q$_{39}$ index calibration by Zinn (1980)
and reported by Harris (1996) (2010 edition) on the modern $ZW$ scale,
[Fe/H]$_{ZW}=-1.77$.
The iron abundance of NGC~6333 has also been estimated by Costar \& Smith (1988) from
the Preston (1959) $\Delta S$ parameter estimated on V1 and V3. These authors
calculated a [Fe/H] value of $-1.93$ and $-1.45$ for these two variables respectively,
for an average of $-1.71$. They used the $\Delta S$-[Fe/H] calibration of Butler
(1975). Had they used the calibration of Suntzeff, Kraft \& Kinman (1994) for RRab
stars or Jurcsik's (1995) or Fernley's et al. (1998) their average [Fe/H] would have
been $-1.72$, $-1.73$ and $-1.92$ respectively. The value [Fe/H]=$-1.72$ is commonly
cited in the literature on NGC~6333, most likely from the $\Delta S$ result. We
have to note however that the $\Delta S$ values were obtained  only on two RR Lyrae
stars (V1 and V3) at a single phase and that the method is strongly phase dependent.
We should also keep in mind that V3 is a clear Blazhko variable. Thus, despite the
good numerical agreement with our results we do not find the comparison of particular
relevance, and believe that our result [Fe/H]$_{ZW}=-1.70\pm0.01$ is more solidly
sustained.

The weighted mean $M_V$ values for the RRab and RRc stars are 0.467$\pm$0.001 mag
and
0.554$\pm$0.001 mag respectively (see Table \ref{fisicos}) and will be used in section
\ref{sec:distance} to estimate the mean distance to the cluster after differential
reddening is considered.

\subsection{RR Lyrae luminosities and effective temperatures}
\label{sec:masses}

The values of $M_V$ in Table~\ref{fisicos} were transformed into
$\log~L/L_\odot = -0.4 (M_V - M_{\rm bol}^{\odot} + BC)$. The bolometric correction
was
calculated using the formula $BC= 0.06$
[Fe/H]$_{ZW} + 0.06$ given by Sandage \& Cacciari (1990). We adopted the value
$M_{bol}^{\odot} = 4.75$ mag.

The effective temperature $T_{\rm eff}$ can be estimated for RRab stars from the
calibrations of Jurcsik (1998):

\begin{equation}
	\log(T_{\rm eff})= 3.9291 ~-~ 0.1112~(V - K)_o ~-~ 0.0032~{\rm [Fe/H]},
\end{equation}

\noindent
with 

$$ (V - K)_o= 1.585 ~+~ 1.257~P ~-~ 0.273~A_1 ~-~ 0.234~\phi^{(s)}_{31} ~+~ $$
\begin{equation}
~~~~~~~ ~+~ 0.062~\phi^{(s)}_{41}.
\end{equation}

For the RRc stars the calibration of Simon \& Clement (1993) can be used:

\begin{equation}
\label{eq:SC93}
	\log(T_{\rm eff}) = 3.7746 ~-~ 0.1452~\log(P) ~+~ 0.0056~\phi^{(c)}_{31}.
\end{equation}

The validity and caveats of the above calibrations have been discussed in several
recent papers (Cacciari et al. 2005; Arellano Ferro et al. 2008; 2010; Bramich et al.
2011) and the reader is referred to them for the details. We list the obtained $T_{\rm
eff}$ values for the RR Lyrae stars in NGC 6333 for comparison with similar work in
other clusters.

\subsection{RR Lyrae masses and radii}
\label{sec:masses}

Once the period, luminosity and temperature are known for each RR Lyrae star, its
mass and radius can be estimated from the equations: $\log~M/M_{\odot} =
16.907 - 1.47~ \log~P_F + 1.24~\log~(L/L_{\odot}) - 5.12~\log~T_{\rm eff}$ (van Albada
\& Baker 1971) and $L$=$4\pi R^2 \sigma T^4$ respectively.
The masses and radii are given in Table \ref{fisicos} in solar
units.

\subsection{Distance to NGC~6333 from the RR Lyrae stars}
\label{sec:distance}

The weighted mean $M_V$, calculated for the RRab and RRc in Table \ref{fisicos} 
can be used to estimate the true distance modulus; $V-M_V = 5 {\log}~d - 5 +
3.1E(B-V)$. The individual colour excesses are listed in Table
\ref{tab:fourier_coeffs}
which were calculated after differential reddening was considered ($\S$
\ref{sec:reddening}). Although the internal
errors in $M_V$ are small, given the mean magnitude dispersion in the HB, a better
estimate of the uncertainty in the distance is the standard deviation of the mean
and so we find the distance modulus of
$14.527\pm0.052$ mag and $14.482\pm 0.081$ mag using the RRab and RRc stars
respectively, which
correspond to the distances $8.04\pm 0.19$ and $7.88\pm 0.30$ kpc.  

The above distances for the RRab and RRc stars are calculated from independent
empirical calibrations, with their own systematic uncertainties, hence
they should be considered as two independent estimates. The correction for
differential reddening  has contributed to the good agreement between these two
estimates of the distance and to reduce the uncertainties.

The distance to NGC~6333 listed in the catalogue of Harris (1996) (2010 edition) is
7.9 kpc, estimated from the mean $M_V$ magnitudes calculated by Clement \& Shelton
(1999). The calibrations of $M_V$ used by Clement \& Shelton for the RRab and RRc
stars are the same as our equations \ref{eq:ḰW01} and \ref{eq:K98} but before
correcting the zero points as discussed in $\S$ \ref{sec:FEMV}. Also, in their
calculation of the distance the differential reddening was not taken into account.
These facts may account for the small difference in the distance we derive for the
cluster.

The most recent discussion on the distance of the cluster is perhaps the one given by
Casetti-Dinescu et al. (2010) in which they adopt the distance 7.9 kpc from Harris
(1996) (2010 edition) but argue that due to reasonable uncertainties in the distance
(~10\%) the alternative distance of 8.6 kpc is selected such that it places the
cluster on the opposite side of the Galactic centre. In our opinion and given our
results, 10\% is large for a distance error and note that if our distance is
adopted then the cluster would be located on the near side of the Galactic centre.

\section{Bailey diagram and Oosterhoff type}

Using the periods listed in Table \ref{variables}, we calculate mean periods of
0.639 and 0.336~d for the 8 RRab and 10 RRc stars, respectively, that are cluster members (i.e.
excluding V13 and V33), and excluding the RRd star V19. These values clearly identify
NGC~6333 as an Oosterhoff type II cluster.

\begin{figure} 
\includegraphics[width=8.0cm,height=12.0cm]{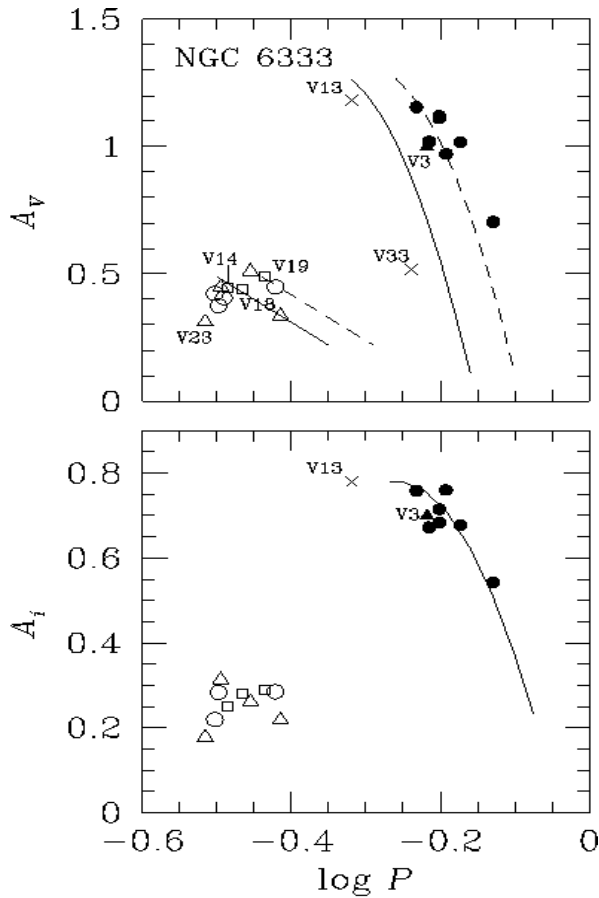}
\caption{The RR Lyrae stars in NGC 6333 on the Amplitude-Period
plane or Bailey diagram for the $V$ and $I$ bands. In the top panel the solid lines
represent the average distributions of
fundamental and first overtone RRL's in M3. The segmented lines are the
loci for the evolved
stars according to  Cacciari et al. (2005). Filled symbols are used for RRab stars and
open symbols for RRc stars. Circles represent stars with no apparent signs of
amplitude modulations and triangles indicate stars with clear and prominent 
amplitude modulations or Blazhko variables. Variables with
secular period change, V14 and V18 and the double mode star V19 are shown as squares.
V13 and V33 are not cluster members. The bottom panel shows the case for the $I$
amplitudes. The solid curve is the locus defined by Arellano Ferro et al. (2011) for
the RRab stars in the OoII cluster NGC 5024.}
\label{fig:Bailey}
\end{figure}

The Bailey diagram (log $P$ vs $A_V$ and log $P$ vs $A_i$) for the RR Lyrae variables
is shown in Fig. \ref{fig:Bailey}.
The RRab stars have longer periods for a given amplitude than their counterparts 
in the OoI cluster M3. This is also seen in the OoII clusters M53 (Arellano Ferro et
al. 2011), M15 and M68 (Cacciari
et al. 2005). This fact confirms the OoII type of NGC~6333.
V13 and V33 are discordant stars and this supports the idea that they are
not cluster members (Clement et al. 1984, $\S$ \ref{sec:IND_STARS}). Other
than V13 and V33 there are no RRab stars with peculiar amplitude, which gives support
to the physical parameters obtained in $\S$ \ref{sec:RRLstars} from the light curve
Fourier decomposition. The RRc stars show some scatter which is likely due to
the amplitude modulations
observed in the population of RRc stars in NGC 6333, the majority of the RRc stars
seem to show some amplitude and/or phase modulations that can be attributed to the
Blazhko effect (see $\S$ \ref{sec:IND_STARS} and Fig.~\ref{VARSc}). A similar case was
found in NGC~5024 which probably contains the largest sample of RRc Blazhko variables
(Arellano Ferro et al. 2011; 2012).

As in Arellano Ferro et al. (2011) for NGC 5024, in the bottom panel of Fig.
\ref{fig:Bailey} we show the Bailey diagram using the amplitudes in the $I$-band, or
$A_i$ in Table \ref{variables}, for NGC 6333. The solid curve is the fit calculated by
Arellano Ferro et al.  for the RRab stars in NGC 5024 (their eq. 3). Being the two
clusters of the OoII type and of similar metallicity, the match is rewarding.

\section{Conclusions}
In their analysis on the completeness of the variable stars sample in NGC 6333,
Clement \& Shelton (1996) concluded that the discovery of
new RRc stars was unlikely but that some RRab stars might have escaped their
attention. In fact,
we have not found any new RR Lyrae, neither RRab nor RRc, in the corresponding field
of Clement \& Shelton's images. However, we found in this work two RRc stars, V22 and
V23, and one RRab, V33; the three of them are relatively isolated in the outskirts of
the cluster. While V22 and V23 are clear members of NGC~6333, V33 is a field RRab
further away than the cluster. Likewise we corroborate that the RRab star V13 is not
a cluster member. We have also been able to find three new eclipsing binaries and
seven long period variables. Pulsation period refinements have been calculated for
nearly all variables. Accurate celestial coordinates and a finding chart for all
previously known and new variables are provided.

Although a cluster membership confirmation from radial velocity data would be
necessary, we argue that V12 is a cluster member since
the correction from interstellar reddening places this star in the cepheid
instability strip at about 1 magnitude above the HB. Although V12 is about 5 arcmin
away from the cluster center, it has been noted by Clement et al. (2001)
in the CVSGC (2012 update) that the star is within the tidal radius of NGC~6333 of $\sim$
8 arcmin. V12 also follows the P-L relation of ACs pulsating in the fundamental mode 
of Pritzl et al. (2002). While ACs are common in dwarf spheroidal galaxies 
(e.g. Pritzl et al. 2002; 2005; Nemec et al.
1994), they are rare in globular clusters; only four are presently known, V19 in NGC
5466 (Zinn \& Dahn 1976) and three in $\omega$ Cen (Kaluzny et al. 1997). We argue
that V12 in NGC 6333 is an AC.

Among the RR Lyrae stars we have identified the double mode or RRd nature of V19 and
the secular period changes in V14 and V18 at the rates of 4.67
and 11.5 d Myr$^{-1}$ respectively. We stress that
similar to NGC 5024 (Arellano Ferro et al. 2012), NGC 6333 has a rather large
number of Blazhko stars among the RRc population.

A deep search for variability among the Blue Stragglers in the cluster was conducted
but none was found. If SX Phe stars do exist in the cluster they must be of
amplitudes smaller than the detection limit of our data.

The Fourier decomposition of the light curves of 9 RRab and 7 RRc stars was performed
and individual values of [Fe/H], $M_V$, $\log~L/L_\odot$, $T_{\rm eff}$ and stellar
mass and radius were calculated using ad hoc semiempirical calibrations.
The weighted mean values of the iron abundance of selected stars gives a cluster mean
metallicity of  [Fe/H]$_{ZW}=-1.70\pm0.01$ in the Zinn \& West (1985)
scale or [Fe/H]$_{UVES}=-1.67\pm0.01$ in the scale defined more recently by Carretta
et al. (2009).  
The weighted mean values of the absolute magnitude of the RRab and the RRc stars lead
to a distance of $8.04\pm 0.19$ and $7.88\pm 0.30$ kpc respectively.
In calculating these distances the heavy differential reddening affecting the cluster
was taken into account by using the detailed reddening map of Alonso-Garc\'ia et al.
(2012).

\section*{Acknowledgments}

We acknowledge an anonymous referee for very relevant input and comments. 
We are grateful to the TAC's of the Hanle and La Silla observatories for generous
telescope time allocation to this project and to the support astronomers of IAO, at
Hanle and CREST (Hosakote) for their very efficient help while acquiring the data.
This project was supported by DGAPA-UNAM grant through project IN104612 and by
the INDO-MEXICAN collaborative program by DST-CONACyT. NK acknowledges an ESO
Fellowship. The research leading to these results has received funding from the
European Community's Seventh Framework Programme (/FP7/2007-2013/) under grant
agreement No 229517. OW (aspirant FRS - FNRS), AE, YD, DR (FRIA PhD student), and J.
Surdej acknowledge support from the Communaut\'{e} fran\c{c}aise de
Belgique -- Actions de recherche concert\'{e}es -- Acad\'{e}mie universitaire
Wallonie-Europe. "
TCH gratefully acknowledges financial support from the Korea Research Council for
Fundamental Science and Technology (KRCF) through the Young Research Scientist
Fellowship Program. TCH acknowledges financial support from KASI (Korea Astronomy and
Space Science Institute) grant number 2012-1-410-02.
KA, DB, MD, MH and CL are supported by NPRP grant NPRP-09-476-1-78 from the Qatar
National Research Fund (a member of Qatar Foundation).
M.R. acknowledges support from FONDECYT postdoctoral fellowship N°3120097.
The Danish 1.54m telescope is operated
based on a grant from the Danish Natural Science Foundation (FNU).
Funding for the Centre for Star and Planet Formation is provided by the Danish
National Research Foundation.
CS received funding from the European Union Seventh Framework Programme
(FP7/2007-2013) under grant agreement no. 268421.
HK acknowledges support from a Marie-Curie Intra-European Fellowship. 

This work has made a large use of the SIMBAD and ADS services.

\noindent
----------------------------------------------------------------------------
\noindent
\\$^{1}$Instituto de Astronom\1a, Universidad Nacional Aut\'onoma de M\'exico.
Ciudad Universitaria CP 04510, Mexico
\\$^{2}$European Southern Observatory, Karl-Schwarzschild-Stra$\beta$e 2, 85748
Garching bei M\"{u}nchen, Germany
\\$^{3}$SUPA, School of Physics and Astronomy, University of St.
Andrews, North Haugh, St Andrews, KY16 9SS, United Kingdom
\\$^{4}$Indian Institute of Astrophysics, Koramangala 560034, Bangalore, India
\\$^{5}$Niels Bohr Institute, University of Copenhagen, Juliane Maries vej 30,
2100 Copenhagen, Denmark \label{nbi}
\\$^{6}$Qatar Foundation, P.O. Box 5825, Doha, Qatar \label{qnrf}
\\$^{7}$Department of Astronomy, Boston University, 725 Commonwealth Ave,
Boston, MA 02215, United States of America \label{bostonu}
\\$^{8}$Dipartimento di Fisica ``E.R Caianiello", Universit di Salerno, Via
Ponte Don Melillo, 84084 Fisciano, Italy \label{salerno}
\\$^{9}$Istituto Nazionale di Fisica Nucleare, Sezione di Napoli, Italy
\label{fisicanucleare}
\\$^{10}$Istituto Internazionale per gli Alti Studi Scientifici (IIASS), Vietri
Sul Mare (SA), Italy \label{iiass}
\\$^{11}$Institut d'Astrophysique et de G\'{e}ophysique, Universit\'e de Li\`ege,
All\'{e}e du 6 Ao\^{u}t
17, Sart Tilman, B\^{a}t.\ B5c, 4000 Li\`ege, Belgium \label{liege}
\\$^{12}$Astronomisches Rechen-Institut, Zentrum f\"{u}r Astronomie der
Universit\"{a}t Heidelberg (ZAH),  M\"{o}nchhofstr.\ 12-14, 69120 Heidelberg, Germany
\label{ari}
\\$^{13}$Institut f\"{u}r Astrophysik, Georg-August-Universit\"{a}t,
Friedrich-Hund-Platz 1, 37077 G\"{o}ttingen, Germany \label{gottingen}
\\$^{14}$Centre for Star and Planet Formation, Geological Museum, {\O}ster
Voldgade 5, 1350 Copenhagen, Denmark \label{copenhagen}
\\$^{15}$Korea Astronomy and Space Science Institute, Daejeon 305-348, Korea
\label{kasi}
\\$^{16}$Jodrell Bank Centre for Astrophysics, University of Manchester, Oxford
Road,Manchester, M13 9PL, UK \label{manchester}
\\$^{17}$Max Planck Institute for Astronomy, K\"onigstuhl 17, 69117 Heidelberg,
Germany \label{mpia}
\\$^{18}$Department of Astronomy, Ohio State University, 140 West 18th Avenue,
Columbus, OH 43210, United States of America \label{osu}
\\$^{19}$Departamento de Astronom\'ia y Astrof\'isica, Pontificia Universidad
Cat\'olica de Chile, Av. Vicu\~na Mackenna 4860, 7820436 Macul,
Santiago, Chile\label{puc}
\\$^{20}$Department of Physics, Sharif University of Technology, P.~O.\ Box
11155--9161, Tehran, Iran \label{sharif}
\\$^{21}$Perimeter Institute for Theoretical Physics, 31 Caroline St. N.,
Waterloo ON, N2L 2Y5, Canada \label{perimeter}
\\$^{22}$INFN, Gruppo Collegato di Salerno, Sezione di Napoli, Italy
\label{infn}
\\$^{23}$Max Planck Institute for Solar System Research, Max-Planck-Str. 2, 37191
Katlenburg-Lindau, Germany \label{mps}
\\$^{24}$Astrophysics Group, Keele University, Staffordshire, ST5 5BG, United
Kingdom \label{keele}
\\$^{25}$Hamburger Sternwarte, Universit\"{a}t Hamburg, Gojenbergsweg 112, 21029
Hamburg
\\$^{26}$Instituto de Astronom\'ia - UNAM, Km 103 Carretera Tijuana Ensenada, 422860,
Ensenada (Baja Cfa), Mexico
\\$^{27}$Main Astronomical Observatory, Academy of Sciences of Ukraine, vul.
Akademika Zabolotnoho 27, 03680 Kyiv, Ukraine
\end{document}